\documentclass[aps,pra,twocolumn,showpacs,superscriptaddress,floatfix,10pt]{revtex4-1}
\usepackage{framed}
\usepackage{graphicx}
\usepackage{amsmath}
\usepackage{epsfig}
\usepackage{helvet}
\usepackage{amssymb}
\usepackage{framed}
\usepackage{bbold}
\usepackage{subfigure}

\newcommand{\ket}[1]{\mbox{$| #1 \rangle$}}

\def\cM{\Lambda}
\allowdisplaybreaks[3]
\def\letter{paper } 
\def\appendix{Appendix}

\begin{document}

\title{
Spacetime symmetries and conformal data \\in the continuous multi-scale entanglement renormalization ansatz}
\author{Q. Hu}
\email{qhu@perimeterinstitute.ca}
\affiliation{Perimeter Institute for Theoretical Physics, Waterloo, Ontario N2L 2Y5, Canada}  \date{\today}
\affiliation{Department of Physics and Astronomy, University of Waterloo, Waterloo, Ontario N2L 3G1,
Canada}
\author{G. Vidal}
\affiliation{Perimeter Institute for Theoretical Physics, Waterloo, Ontario N2L 2Y5, Canada}  \date{\today}

\begin{abstract}
The generalization of the multi-scale entanglement renormalization ansatz (MERA) to continuous systems, or cMERA [Haegeman et al., Phys. Rev. Lett, 110, 100402 (2013)], is expected to become a powerful variational ansatz for the ground state of strongly interacting quantum field theories. In this \letter  we investigate, in the simpler context of Gaussian cMERA for free theories, the extent to which the cMERA state $\ket{\Psi^{\cM}}$ with finite UV cut-off $\Lambda$ can capture the spacetime symmetries of the ground state $\ket{\Psi}$. For a free boson conformal field theory (CFT) in 1+1 dimensions as a concrete example, we build a quasi-local unitary transformation $V$ that maps $\ket{\Psi}$ into $\ket{\Psi^{\cM}}$ and show two main results. (i) Any spacetime symmetry of the ground state $\ket{\Psi}$ is also mapped by $V$ into a spacetime symmetry of the cMERA $\ket{\Psi^{\cM}}$. However, while in the CFT the stress-energy tensor $T_{\mu\nu}(x)$ (in terms of which all the spacetime symmetry generators are expressed) is local, the corresponding cMERA stress-energy tensor $T_{\mu\nu}^{\cM}(x) = V T_{\mu\nu}(x) V^{\dagger}$ is quasi-local. (ii) From the cMERA, we can extract quasi-local scaling operators $O^{\cM}_{\alpha}(x)$ characterized by the exact same scaling dimensions $\Delta_{\alpha}$, conformal spins $s_{\alpha}$, operator product expansion coefficients $C_{\alpha\beta\gamma}$, and central charge $c$ as the original CFT. Finally, we argue that these results should also apply to interacting theories.

\end{abstract}


\maketitle

The study and numerical simulation of interacting quantum many-body systems is an extremely challenging task. Making progress often requires the use of a simplifying variational ansatz, such as the multi-scale entanglement renormalization ansatz (MERA) \cite{MERA}, which aims to describe the ground state of lattice Hamiltonians. The MERA can be visualized as the result of a unitary evolution, running from large distances to short distances, that maps an initial unentangled state into a complex many-body wavefunction by gradually introducing entanglement into the system, scale by scale. The success of the MERA in a large class of lattice systems, including systems with topological order \cite{topoMERA} or at a quantum critical point \cite{MERA,Giovannetti08,Pfeifer09,critMERA}, teaches us that this \textit{entangling evolution in scale} picture is a valid --and computationally powerful!-- way of thinking about ground states and their intricate structure of correlations. With a built-in notion of the renormalization group \cite{RG}, MERA is also actively investigated in several other contexts, from holography \cite{Swingle,HologOthers,Czech} (as a discrete realization of the AdS/CFT correspondence \cite{Maldacena}) to statistical mechanics \cite{TNR}, error correction \cite{ErrorCorrection}, and machine learning \cite{MachineLearning}. 

The MERA formalism can also be applied to a quantum field theory (QFT), after introducing a lattice as a UV regulator. For instance, when applied to a conformal field theory (CFT) \cite{BPZ,Cardy,CFTbook}, corresponding to a critical QFT, lattice MERA has been seen to accurately reproduce the universal properties of the corresponding quantum phase transition (as given by the conformal data) \cite{Pfeifer09,critMERA}. However, introducing a lattice has a devastating effect on the spacetime symmetries of the original QFT, with e.g. translation and rotation invariance being reduced to invariance under a discrete subset of translations and rotations. To overcome this difficulty, Haegeman, Osborne, Verschelde, and Verstraete \cite{cMERA} proposed the continuous MERA (cMERA), which describes an entangling evolution of the quantum field degrees of freedom, from some IR length scale large all the way down to a UV length scale $1/\Lambda$, directly in the continuum, that is, without introducing a lattice. In this case, the entangling evolution in scale is generated by a Hermitian operator $L+K$ that explicitly preserves translation and rotation invariance. While a fully general cMERA algorithm for interacting QFTs (the truly interesting but much more challenging scenario) is still missing (see however \cite{cMERAint}), the simplified Gaussian version of cMERA, also proposed in Ref. \cite{cMERA}, already provides a valuable proof of principle that lattice MERA can be successfully extended to the continuum, one that has attracted considerable attention in the context of holography \cite{cMERAholog} and can extract non-perturbative information of interacting QFTs \cite{cMERAint}.

In this \letter we explore to what extent, and in which sense, cMERA can preserve the spacetime symmetries of the original QFT. Our starting point is the simple observation that, by construction, a successful cMERA approximation $\ket{\Psi^{\cM}}$ should reproduce the targeted QFT ground state $\ket{\Psi}$ at all length scale all the way down to $1/\Lambda$ (the scale at which the entangling evolution ends). Accordingly, there should exist a quasi-local unitary transformation $V$, acting non-trivially only at short distances $\lesssim 1/\Lambda$, that maps $\ket{\Psi}$ into $\ket{\Psi^{\cM}}$, i.e. $\ket{\Psi^{\cM}} = V \ket{\Psi}$. If this was indeed the case, then we could use $V$ to map any generator $G$ of a symmetry of the ground state, satisfying $G\ket{\Psi}=0$, into a quasi-local version $G^{\cM}\equiv V G V^{\dagger}$ satisfying $G^{\cM}\ket{\Psi^{\cM}}=0$. That is, all symmetries of $\ket{\Psi}$, including its spacetime sysmmetries, would automatically turn into symmetries of $\ket{\Psi^{\cM}}$, which would however be realized quasi-locally. For the sake of concreteness, in this \letter we will formalise the above intuition and explore its implications for a specific QFT, namely the 1+1 free boson CFT, whose spacetime symmetries are given by the conformal group. However, the above result can be seen to hold more generally for any optimized Gaussian cMERA in Ref. \cite{cMERA}, and we expect it to be correct also in the interacting case. We will first show that the optimized cMERA $\ket{\Psi^{\cM}}$ for the 1+1 free boson CFT, as provided in \cite{cMERA}, is invariant under (a quasi-locally generated version of) the global conformal group, which includes scale transformations. We will then see that the quasi-local generator $D^{\cM} \equiv VDV^{\dagger}$ (where $D$ is the generator of scale transformations or dilations in the CFT) is equal to the generator $L+K$ of the \textit{entangling evolution in scale} that defined the cMERA in the first place. Finally, as a practical application, we will explain how the (exact!) conformal data of the target CFT can be extracted from $\ket{\Psi^{\cM}}$ by studying the set of smeared scaling operators $O^{\cM}_{\alpha}(x)$ associated to $D^{\cM} = L+K$, thus establishing that cMERA can capture the universality class of a quantum phase transition. 

\textit{Continuous MERA for a massless free boson.---} Consider the 1+1 dimensional massless Klein Gordon QFT,
\begin{equation}
H = \frac{1}{2} \int_{-\infty}^{\infty} dx ~:\!\left[\pi(x)^2+(\partial\phi(x))^2\right]\!:
\label{eq:boson_CFT1}
\end{equation}
for bosonic conjugate field operators $\phi(x)$ and $\pi(x)$, with $[\phi(x),\pi(y)] = i\delta(x-y)$. $H$ can be diagonalized \cite{Peskin},
\begin{eqnarray}
H &=&\frac{1}{2} \int dk~:\!\left[\pi(-k)\pi(k) + k^2 \phi(-k)\phi(k)\right]\!: \label{eq:boson_CFT2}\\
&=&\int dk~ |k|~ a(k)^{\dagger} a(k).
\label{eq:boson_CFT3}
\end{eqnarray}
by first introducing Fourier space mode operators $\phi(k) \equiv \frac{1}{\sqrt{2\pi}}\int dx~ e^{-ikx} \phi(x)$ and $\pi(k) \equiv \frac{1}{\sqrt{2\pi}}\int dx~ e^{-ikx} \pi(x)$ 
and then annihilation operators $a(k)$,
\begin{eqnarray} \label{eq:aCFT}
a(k) &\equiv& \sqrt{\frac{|k|}{2}}\phi(k) + i  \sqrt{\frac{1}{2|k|}}\pi(k),
\end{eqnarray}
with $[a(k),a(q)^{\dagger}] = \delta(k-q)$. Above, the normal ordering $:\!\!A\!\!:$ of an operator $A$ is defined as usual by placing the $a$'s to the right of the $a^{\dagger}$'s [e.g., if $A = a(k)a(q)^{\dagger}$, then $:\!\!A\!\!: ~= a(q)^{\dagger} a(k)$]
and ensures a vanishing energy for the ground state $\ket{\Psi}$ of $H$, which is characterized by the infinite set of linear constraints
\begin{equation} \label{eq:CFT}
a(k)\ket{\Psi}=0, ~~~\forall k.
\end{equation}

\begin{figure}[t]
\centering
\includegraphics[width=8.5cm]{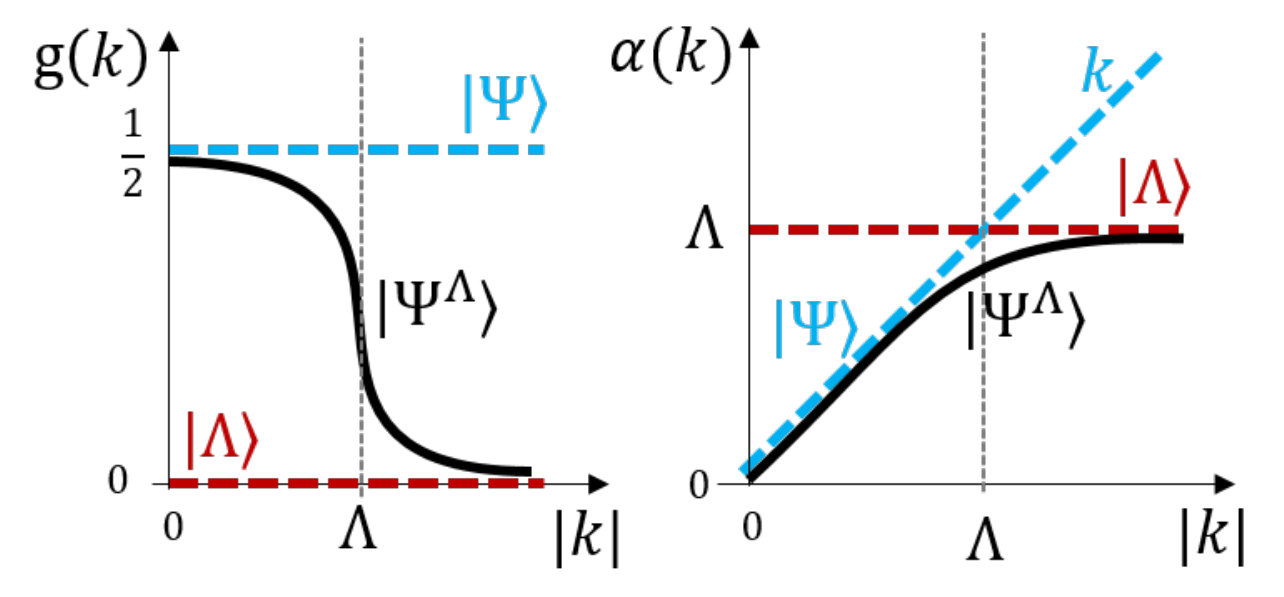}
\caption{
Left: function $g(k) = \exp\left(-e^{-\gamma}(k/\Lambda)^2\right)/2$ for the optimized Gaussian cMERA of Ref. \cite{cMERA}, where $\gamma \approx 0.57722$ is Euler's constant (see \appendix). For comparison, the CFT dilation operator $D$ such that $D\ket{\Psi}=0$ corresponds to choosing $g(k)=1/2$ and the non-relativistic dilation operator $L$ such that $L\ket{\Lambda}=0$ corresponds to $g(k)=0$). 
Right: function $\alpha(k)$ for $g(k)$, interpolating between the linear dependence $k$ for the CFT ground state $\ket{\Psi}$  at small $k$ and the constant value $\Lambda$ for the product state $\ket{\Lambda}$ at large $k$. 
}
\label{fig:Boson}
\end{figure}

On the other hand, the Gaussian cMERA $\ket{\Psi^{\cM}}$ for this CFT, as proposed and optimized in Ref. \cite{cMERA}, reads
\begin{equation} \label{eq:cMERAdef}
\ket{\Psi^{\cM}} \equiv U(0,-\infty) \ket{\Lambda},
\end{equation}
namely it is the result of applying a unitary evolution $U$ to a product (unentangled) state $\ket{\Lambda}$, characterized by
\begin{equation} \label{eq:prod}
\left(\sqrt{\frac \Lambda 2} \phi(k)+\frac{i}{\sqrt{2\Lambda}} \pi(k) \right)\ket{\Lambda} =0,~~\forall k.
\end{equation}
In the context of a scale invariant QFT, $U$ reads 
\begin{equation} \label{eq:U}
U(s_{UV},s_{IR}) \equiv e^{-i\left(L+K\right)(s_{UV}-s_{IR})},
\end{equation}
where the generator of non-relativistic dilations $L$ and the so-called entangler $K$ are given by 
\begin{eqnarray}
L &\equiv& \frac 1 2 \int dk \Big[  \pi(-k) (k\partial_k+\frac 1 2) \phi(k)+ h.c. \Big],\\
K &\equiv& \frac 1 2 \int dk~g(k) \big[  \pi(-k)\phi(k)+h.c. \big],
\end{eqnarray}
and the optimized function $g(k)$ in Fig. \ref{fig:Boson} smoothly approaches $1/2$ and $0$ for small and large $k$, respectively,
\begin{equation} \label{eq:g}
  g(k) \sim \left\{ \begin{array}{cl}
  1/2,&  ~~~~|k|\ll \Lambda,\\
  0,& ~~~~ |k| \gg \Lambda.\\
  \end{array}\right.
\end{equation}

By introducing new annihilation operators $a^{\cM}(k)$,
\begin{equation}\label{eq:a_cM}
a^{\cM}(k) \equiv \sqrt{\frac{\alpha(k)}{2}}\phi(k) + i  \sqrt{\frac{1}{2\alpha(k)}}\pi(k),
\end{equation}
with $[a^{\cM}(k),a^{\cM}(q)^{\dagger}] = \delta(k-q)$, here we start by pointing out that the cMERA state in Eq. \ref{eq:cMERAdef} can be equivalently specified by the modified set of linear constraints
\begin{equation} \label{eq:cMERA}
a^{\cM}(k)\ket{\Psi^{\cM}} = 0, ~~~\forall k,
\end{equation}
provided $\alpha(k)$ and $g(k)$ are related by $d \alpha(k)/d k=2 g(k)\alpha(k)/k$
(see \appendix). For the $g(k)$ in Fig. \ref{fig:Boson}, this implies 
\begin{equation} \label{eq:alpha}
  \alpha(k) \sim \left\{ \begin{array}{cl}
  |k|,&  ~~~~|k|\ll \Lambda~~~\mbox{(CFT limit)},\\
  \Lambda,& ~~~~ |k| \gg \Lambda~~~\mbox{(product state limit)}.
  \end{array}\right.
\end{equation}  
To gain insight into the structure of $\ket{\Psi^{\cM}}$, we notice that the constraints it satisfies (Eqs. \ref{eq:a_cM}-\ref{eq:alpha}) interpolate between the constraints of the CFT ground state $\ket{\Psi}$ (Eqs. \ref{eq:aCFT}-\ref{eq:CFT}) at small $k$ and those of the product state $\ket{\Lambda}$ (Eq. \ref{eq:prod}) at large $k$. In other words, the optimized cMERA should somehow behave as the CFT ground state $\ket{\Psi}$ at large distances $x\gg 1/\Lambda$ and as the product state $\ket{\Lambda}$ at short distances $x \ll 1/\Lambda$ \cite{cMERA}. A direct calculation \cite{Franco} confirms that, in sharp contrast to the target CFT, in cMERA correlation functions and entanglement entropy remain finite/regulated at short distances, implying that $\ket{\Psi^{\cM}}$ has a built-in UV cut-off.

\textit{Smearing symplectic transformation.---} In order to investigate the spacetime symmetries of $\ket{\Psi^{\cM}}$, let us introduce the unitary map $V$, defined by
\begin{eqnarray}
V\phi(k)V^{\dagger} &=& \sqrt{\frac{\alpha(k)}{|k|}}\phi(k) \equiv \phi^{\cM}(k), \label{eq:symp1}\\
V\pi(k)V^{\dagger} &=&  \sqrt{\frac{|k|}{\alpha(k)}}\pi(k) \equiv \pi^{\cM}(k). \label{eq:symp2}
\end{eqnarray}
This map implements a symplectic transformation (preserving canonical commutation relations) that transforms $a(k)$ into $Va(k)V^{\dagger} = a^{\cM}(k)$ and therefore the CFT ground state into the cMERA, V \ket{\Psi} = \ket{\Psi^{\cM}}. The transformation is quasi-local: $V$ maps the sharp field operators $\phi(x)$ and $\pi(x)$ into operators $\phi^{\cM}(x)$ and $\pi^{\cM}(x)$ that are smeared over a length $1/\Lambda$. Indeed, through a Fourier transform of $\phi^{\cM}(k)$ and $\pi^{\cM}(k)$ we obtain
\begin{eqnarray}
\phi^{\cM}(x) = V\phi(x)V^{\dagger} = \int dy~  \mu_\phi(x-y) \phi(y),\\
\pi^{\cM}(x) = V\pi(x)V^{\dagger} = \int dy~  \mu_\pi(x-y) \pi(y),
\end{eqnarray}
where $\mu_\phi$ and $\mu_\pi$ are distributional Fourier transforms \cite{Hadamard} of $\sqrt{\frac{\alpha(k)}{|k|}}$ and $\sqrt{\frac{|k|}{\alpha(k)}}$ that are upper bounded by an exponentially decaying function for $\Lambda|x|\gg 1$ (see \appendix). We also note that since $V$ acts diagonally in momentum space, its action by conjugation commutes with the spatial derivative $\partial_x$. For instance, 
\begin{equation}
(\partial_x \phi)^{\cM}(x) \equiv V \partial_x\phi(x) V^{\dagger} = \partial_x \left(V\phi(x)V^{\dagger}\right) = \partial_x(\phi^{\cM}(x)), \nonumber
 \end{equation} 
with the smearing function $\mu_{\phi'}(y)$ for $(\partial_x \phi)^{\cM}(x)$ being the derivative of the smearing function for $\phi^{\cM}(x)$, that is $\mu_{\phi'}(y) = d\mu_{\phi}(y)/dy = 1/\sqrt{2\pi}\int dk~e^{iky}\left(ik\right)  \sqrt{\frac{\alpha(k)}{|k|}}$.
In particular, the right moving and left moving fields $\partial \phi (x) \equiv (\partial_x \phi(x) - \pi(x))/2$  and   $\bar{\partial} \phi (x) \equiv (\partial_x \phi(x) + \pi(x))/2$ of the CFT \cite{Euclidean} are mapped into smeared right and left moving fields, e.g.
\begin{equation} \label{eq:cpartial}
\left(\partial \phi\right)^{\cM} (x) \equiv V\partial \phi(x) V^{\dagger} = \frac{1}{2}\left(\partial_x \phi^{\Lambda}(x) - \pi^{\cM}(x)\right).
\end{equation}

\textit{Quasi-local stress-energy tensor.---} The spacetime symmetry generators of the CFT are naturally expressed in terms of the symmetric, traceless stress-energy tensor $T_{\mu\nu}(x)$ \cite{CFTbook}, with components $T_{00}(x)   =  :\!\!\left[\pi(x)^2 + (\partial_x \phi(x))^2\right]\!\!:/2  \equiv h(x)$ and $T_{01} (x)= -:\!\!\pi(x)\partial_x\phi(x)\!\!: \equiv p(x)$, where $h(x)$ and $p(x)$ are the energy and momentum densities. In close analogy, the quasi-local stress-energy tensor $T^{\cM}_{\mu\nu}(x) \equiv V T_{\mu\nu}(x) V^{\dagger}$ defines quasi-local energy and momentum densities,
\begin{eqnarray}
h^{\cM}(x) &\equiv& T_{00}^{\cM}(x) = \frac{1}{2}:\!\left[\pi^{\cM}(x)^2 + (\partial_x \phi^{\cM}(x))^2\right]\!:,\\
p^{\cM}(x) &\equiv& -iT_{01}^{\cM}(x) = -:\! \pi^{\cM}(x) \partial_x\phi^{\cM}(x)\!:,
\end{eqnarray}
where we have used that for any two operators $A(x)$ and $B(x)$, $(A(x)B(y))^{\cM} \equiv V\left( A(x) B(y) \right)V^{\dagger} = \left(V A(x)V^{\dagger} \right) \left(V B(y)V^{\dagger} \right) = A^{\cM}(x)B^{\cM}(y) $ and the normal order is now with respect to the annihilation operators $a^{\cM}(k)$ in Eq. \ref{eq:a_cM}. As argued earlier, any generator of a symmetry of the CFT ground state $\ket{\Psi}$ is mapped into a quasi-local generator of a symmetry of $\ket{\Psi^{\cM}}$. Let us now elaborate this point with a few explicitly examples.

\textit{Translations in time and space.---} Hamiltonian $H = \int dx ~h(x)$ in Eq. \ref{eq:boson_CFT1} and the momentum operator $P \equiv \int dx~p(x)$, generators of translations in time $(t,x) \rightarrow (t+t_0,x)$ and in space $(t,x) \rightarrow (t,x+x_0)$, are mapped into 
\begin{eqnarray}
H^{\cM} &=& \int dx~ h^{\cM}(x) = \int dk~ |k| ~a^{\cM}(k)^{\dagger}a^{\cM}(k), \\
P^{\cM} &=& \int dx~p^{\cM}(x) = \int dk~ k ~a^{\cM}(k)^{\dagger}a^{\cM}(k),
\end{eqnarray}
whose expressions in terms of the annihilation operators $a^{\cM}(k)$ make manifest that $H^{\cM} \ket{\Psi^{\cM}} = 0$, $P^{\cM} \ket{\Psi^{\cM}} =0$, and that $\ket{\Psi^{\cM}}$ is the ground state of the quasi-local Hamiltonian $H^{\cM}$, since $H^{\cM}\geq 0$. Direct inspection shows that $P^{\cM}=P$. We have thus recovered the known invariance of $\ket{\Psi^{\cM}}$ under space translations generated by $P$ \cite{cMERA} (equivalently, $P^{\cM}$), and have in addition shown its invariance under time translations generated by $H^{\cM}$. Notice, moreover, that a complete set of simultaneous eigenstates of $H^{\cM}$ and $P^{\cM}$, can now be built by applying the creation operators $a^{\cM}(k)^{\dagger}$'s on $\ket{\Psi^{\cM}}$.

\textit{Lorentz boosts and scale transformations.---} We can similarly define cMERA analogues of the generators $B \equiv \int dx~x~h(x)$ and $D \equiv \int dx~x~p(x)$ of boosts $(x,t)\rightarrow \gamma (x-vt,t-vx)$ [where $\gamma \equiv 1/\sqrt{1-v^2}$ is the Lorentz factor and $v$ is the relative velocity] and dilations $(t,x)\rightarrow (\lambda t, \lambda x)$ [where $\lambda$ is the re-scaling factor], namely
\begin{eqnarray}
B^{\cM} &=&  \int dx~x~h^{\cM}(x)\\
&=& i\int_{-\infty}^{\infty} dk~a^{\cM}(k)^{\dagger}sgn(k)\left(k\partial_k + \frac{1}{2}\right)a^{\cM}(k),\\
D^{\cM} &=& \int dx~x~p^{\cM}(x) \\
&=& i\int_{-\infty}^{\infty} dk  ~a^{\cM}(k)^{\dagger}\left(k\partial_k + \frac{1}{2}\right)a^{\cM}(k),
\end{eqnarray}
which again manifestly annihilate $\ket{\Psi^{\cM}}$. Operator $B^{\cM}$ generates a continuous symmetry of $\ket{\Psi^{\cM}}$ related to relativistic invariance and with no counter-part on the lattice. Importantly, a direct computation (see \appendix) shows that $D^{\cM} = L+K$, so that the generator of scale transformations $D^{\cM}$ coincides with the generator $L+K$ of the unitary evolution in scale that defines the cMERA $\ket{\Psi^{\cM}}$. Accordingly, the cMERA state $\ket{\Psi^{\cM}}$ is scale invariant, in spite of containing no entanglement at distances smaller than $1/\Lambda$, if we agree to regard $D^{\cM}= L+K$ as the generator of dilations. The dilations generated by $D^{\cM}$ not only re-scale spacetime in the usual sense, but also introduce or remove entanglement as needed in order to reset the UV cut-off back to $1/\Lambda$. 

We emphasize that, by construction, the operators $H^{\cM}$, $P^{\cM}$, $B^{\cM}$ and $D^{\cM}$ inherit the commutation relations of the CFT generators ($[H^{\cM}, P^{\cM}] =[B^{\cM}, D^{\cM}]=0$, $-i[D^{\cM}, H^{\cM}] = H^{\cM}$, etc) and therefore close the same algebra, which can be extended to the global conformal group and even to the full Virasoro algebra (see \appendix). Thus, the cMERA realizes a quasi-local, smeared version of conformal symmetry. 

In Ref. \cite{cMERA}, Haegeman et al. pointed out that the state $\ket{\Psi^{\cM}}$ recovers scale invariance in the limit $\Lambda \rightarrow \infty$, where it coincides with the target CFT ground state $\ket{\Psi}$. Here we have just argued, in sharp contrast, that $\ket{\Psi^{\cM}}$ is already scale invariant at \textit{finite} $\Lambda$, provided that we adopt $D^{\cM}=L+K$ as the generator of scale transformations. Admittedly, the scale invariance of $\ket{\Psi^{\cM}}$ is a tautology (because $\ket{\Psi^{\cM}}$ had been introduced in Eqs. \ref{eq:cMERAdef}-\ref{eq:U} as a fixed-point of $L+K$!). To see why these unorthodox notions of scale transformation and scale invariance are nevertheless very useful, next we show that they lead to smeared versions of the scaling operators of the theory from which the conformal data of the target CFT can be extracted.


\textit{Quasi-local scaling operators and conformal data.---} Let us thus search for the quasi-local scaling operators $O^{\cM}_{\alpha}(x)$ that transform covariantly under $D^{\cM}$ and $B^{\cM}$, that is, such that (choosing $x=0$ for simplicity)
\begin{eqnarray}
-i[D^{\cM},O^{\cM}_{\alpha}(0)] &=& \Delta_{\alpha}O^{\cM}_{\alpha}(0),\label{eq:DO}\\
-i[B^{\cM},O^{\cM}_{\alpha}(0)] &=& s_{\alpha} O^{\cM}_{\alpha}(0),\label{eq:BO}
\end{eqnarray}
where $\Delta_{\alpha}$ and $s_{\alpha}$ are the scaling dimension and conformal spin of $O^{\cM}_{\alpha}(x)$, respectively \cite{compareJutho}. One could determine $O^{\cM}_{\alpha}$ by solving Eqs. \ref{eq:DO}-\ref{eq:BO}, but there is no need. Indeed, we can instead use $V$ to translate the sharp scaling operators of the CFT (which are already known \cite{CFTbook}) into smeared cMERA scaling operators. A first example is a linear scaling operator of the CFT, namely the right moving field $\partial \phi(x)$ discussed before, which satisfies $-i[D,\partial\phi(0)] = \partial \phi(0)$ and $-i[B,\partial\phi(0)] = \partial \phi(0)$, implying a scaling dimension $\Delta_{\partial\phi} = 1$ and conformal spin $s_{\partial\phi} = 1$. Using the symplectic map $V$ we readily obtain corresponding expressions for $\partial \phi^{\cM}(x)$ in Eq. \ref{eq:cpartial}, namely
\begin{eqnarray}
 -i[D^{\cM},\partial\phi^{\cM}(0)] &=& \partial \phi^{\cM}(0),\\
 -i[B^{\cM},\partial\phi^{\cM}(0)] &=& \partial \phi^{\cM}(0),
\end{eqnarray}
and thus $\partial \phi^{\cM}(x)$ has the same scaling dimension $\Delta_{\partial\phi^{\Lambda}} = 1$ and conformal spin $s_{\partial\phi^{\Lambda}} =1$ by $D^{\cM}$ and $B^{\cM}$ as $\partial \phi(x)$ has by $D$ and $B$. This results extends to \textit{all} scaling operators of the CFT, including e.g. quadratic scaling operators such as the right moving part of the stress-energy tensor, $T(x) =  -2\pi:\partial \phi(x) \partial \phi(x): $, with $\Delta_T= s_T =  2$, or vertex operators $V_{\nu}(x) \equiv ~:\!e^{i\nu \phi(x)}\!:$. Moreover, operators $O^{\cM}_{\alpha}(x)$ also inherit from the CFT its operator product expansion (OPE) coefficients $C_{\alpha\beta\gamma}$. For instance, the OPE $\partial \phi(x) \partial \phi(y) \sim -1/\left(4\pi (x-y)^{2}\right)$, which implies $C_{\partial \phi \partial\phi \mathbb{1}} = -1/4\pi$ (see \appendix), translates into
\begin{equation}
\partial \phi^{\cM}(x) \partial \phi^{\cM}(y) \sim \frac{-1}{4\pi}\frac{1}{(x-y)^{2}},
\end{equation}
and thus identical OPE coefficient. Finally, the central charge $c$ can be obtained from (a translation of) the standard OPE of $T(x)$ with itself, namely
\begin{equation}
T^{\cM}(x)T^{\cM}(y) \sim \frac{c/2}{(x-y)^4} + \frac{2T^{\cM}(y)}{(x-y)^2} + \frac{\partial_y T^{\cM}(y)}{(x-y)},
\end{equation}
which results in $c=1$. 

\textit{Discussion.---} We have seen that the Gaussian cMERA for a 1+1 free boson CFT, as proposed and optimized in Ref. \cite{cMERA}, inherits (a quasi-locally realized version of) the spacetime symmetries of the conformal theory. This result was based on identifying the quasi-local unitary transformation $V$ that maps the CFT ground state $\ket{\Psi}$ into the cMERA $\ket{\Psi^{\cM}}$, and then using it to also map the symmetry generators of the original theory. As an application, we have shown that from the generators $D^{\cM}=L+K$ and $B^{\cM} = \int dx~ x~h^{\cM}(x)$ we can reconstruct all the conformal data of the original CFT, namely the central charge $c$, and the scaling dimensions $\Delta_{\alpha}$ and conformal spins $s_{\alpha}$ of the primary fields, together with their OPE coefficients $C_{\alpha\beta\gamma}$. A similar transformation $V$ can also be built for the optimized Gaussian cMERA of any free QFT analysed in Ref. \cite{cMERA}, including higher dimensional CFTs (invariant under the global conformal group) and massive relativistic QFTs (invariant only under the Poincare group, with a scale dependent entangler $K(s)$). 


We conclude by briefly commenting  on the (non-Gaussian) cMERA for interacting QFTs, for which no optimization algorithm is yet known. Based on the success of MERA \cite{MERA,Pfeifer09,critMERA} for interacting theories on the lattice over the last 10 years, it is reasonable to speculate that a putative interacting cMERA algorithm will produce an optimized state $\ket{\Psi^{\cM}}$ that will again only differ significantly from its target ground state $\ket{\Psi}$ at short distances. Accordingly, a quasi-local unitary $V$ should also exist relating $\ket{\Psi}$ and $\ket{\Psi^{\cM}}$ that maps the generators of symmetries into quasi-local generators. In this way, for instance, we once again expect to be able to extract an accurate estimate of the conformal data of an interacting CFTs from an optimized non-Gaussian cMERA approximation $\ket{\Psi^{\cM}}$.

We thank John Cardy, Bartlomiej Czech, Adrian Franco, Davide Gaiotto, Lampros Lamprou, Rob Myers, Nick Van den Broeck, Pedro Vieira for discussions and comments.
The authors acknowledge support from by NSERC (Discovery grant) and the Simons Foundation (Many Electron Collaboration).  Research at Perimeter Institute is supported by the Government of Canada through Industry Canada and by the Province of Ontario through the Ministry of Research and Innovation.

\section{Appendix A: An entangling evolution in scale}

In this appendix we consider in more detail the Gaussian cMERA $\ket{\Psi^{\cM}}$ for the ground state $\ket{\Psi}$ of the 1+1 free boson CFT discussed in the main text. This specific instance of cMERA was originally proposed and extensively discussed by Haegeman, Osborne, Verschelde, and Verstraete in the supplementary material of Ref. \cite{cMERA}. The goal of this appendix is to provide further details on the new results presented in the main text of our paper. These include an alternative characterization of $\ket{\Psi^{\cM}}$ in terms of a set of annihilation operators, regarding the $L+K$ as a generator $D^{\cM}$ of dilations at finite UV cut-off, and establishing the quasi-local character of the associated scaling operators.

We start by providing some intuition on the first characterization of the cMERA $\ket{\Psi^{\cM}}$ for a CFT. This corresponds to the original definition of cMERA in  \cite{cMERA} specialized to a CFT,
\begin{equation} \label{eqApp:cMERAlim}
\ket{\Psi^{\cM}} = \lim_{s_{IR} \rightarrow \infty} e^{-i (L+K) (0-s_{IR})}\ket{\Lambda},
\end{equation}
which regards $\ket{\Psi^{\cM}}$ as the fixed point of an infinitely long unitary evolution generated by the (scale independent) operator $L+K$, where 
\begin{eqnarray}
L &\equiv& \frac 1 2 \int dk \Big[  \pi(-k) (k\partial_k+\frac 1 2) \phi(k)+ h.c. \Big], \label{eqApp:L}\\
K &\equiv& \frac 1 2 \int dk~g(k) \big[  \pi(-k)\phi(k)+h.c. \big], \label{eqApp:K}
\end{eqnarray}
acting on an initial product state $\ket{\Lambda}$.
We then establish the equivalence between this characterization and a second, alternative characterization in terms of a complete set of annihilation operators $a^{\cM}(k)$, such that $a^{\cM}(k)\ket{\Psi^{\cM}} = 0$ for all $k$. Finally, we show that the symplectic transformation $V$ that maps the target ground state $\ket{\Psi}$ into the cMERA $\ket{\Psi^{\cM}}$, $\ket{\Psi^{\cM}} = V \ket{\Psi}$, is quasi-local, by showing that its action on the sharp field operators $\phi(x)$ and $\pi(x)$ produces quasi-local field operators $\phi^{\cM}(x)$ and $\pi^{\cM}(x)$.

\subsection{Simplified cMERA $\ket{\tilde{\Psi}^{\cM}}$ with sharp cut-off}

\begin{figure}[t]
\centering
\includegraphics[width=8.5cm]{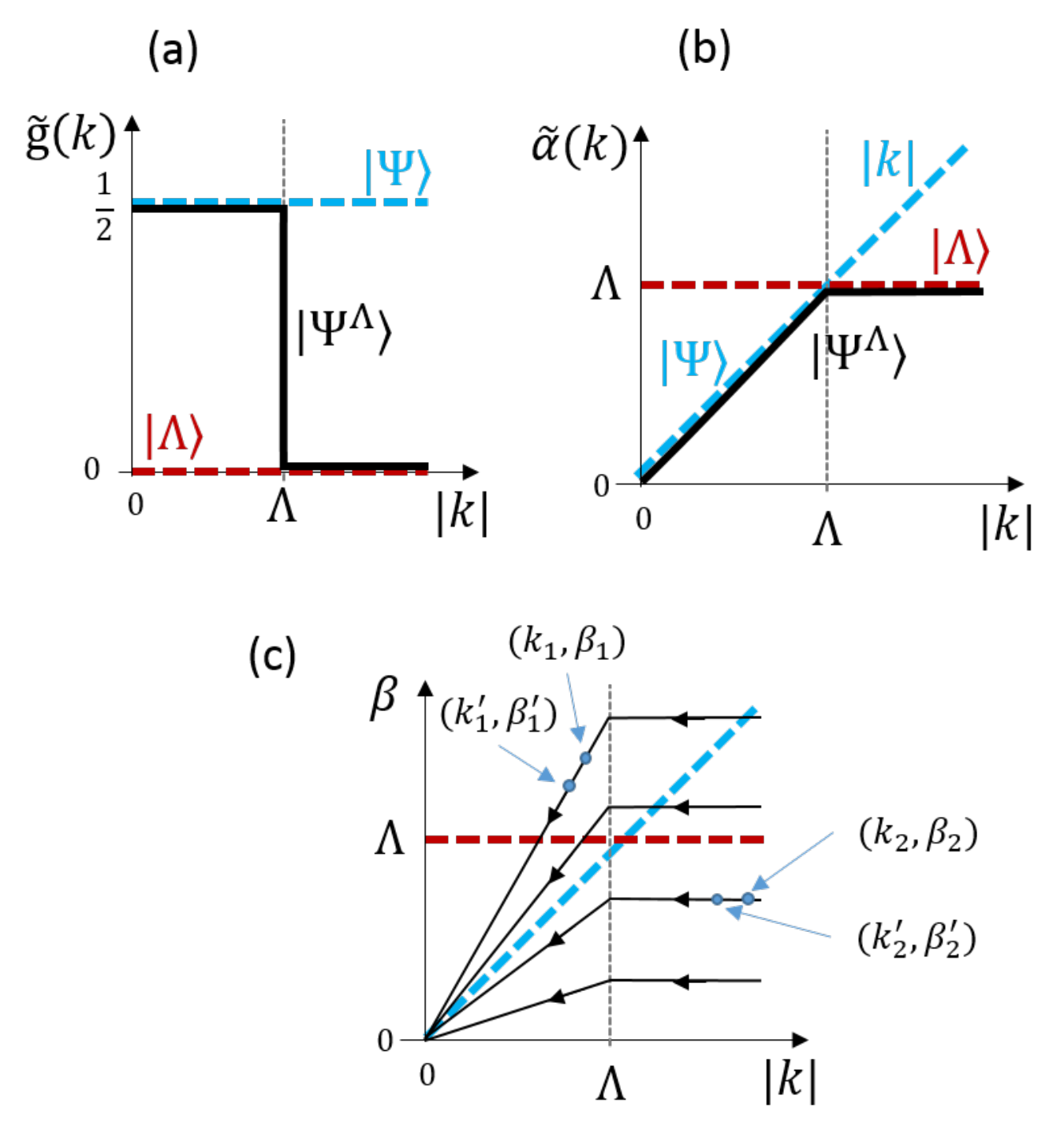}
\caption{
(a) Simplified function $\tilde{g}(k)$ of Eq. \ref{eqApp:g} characterized by a sharp cut-off at $|k| = \Lambda$.
(b) Corresponding simplified $\tilde{\alpha}(k)$ of Eq. \ref{eqApp:alpha_tilde} characterizing the simplified cMERA state $\ket{\tilde{\Psi}^{\cM}}$ of Eq. \ref{eqApp:cMERA_tilde}. 
(c) Flow in the space $(k,\beta)$ of linear constraints described in Eq. \ref{eqApp:constraint1} generated by the simplified $L+\tilde{K}$. Under a small evolution by $L+\tilde{K}$, $(k,\beta)$ becomes $(k',\beta')$, where $\beta'<\beta$ if $|k|\leq \Lambda$ (as illustrated by the $\beta_1$ and $\beta_1'$) and $\beta' = \beta$ if $|k|>\Lambda$ (as illustrated by $\beta_2$ and $\beta_2'$).
}
\label{fig:sharp}
\end{figure}

To gain intuition on the fixed-point characterization of cMERA, we will temporarily work with a simplified choice of $g(k)$ in Eq. \ref{eqApp:K}, denoted $\tilde{g}(k)$,
\begin{equation}  \label{eqApp:g}
  \tilde{g}(k) \equiv \left\{ \begin{array}{cl}
  1/2,&  ~~~~|k| \leq \Lambda,\\
  0,& ~~~~ |k| >\Lambda.\\
  \end{array}\right.
\end{equation} 
corresponding to a sharp cut-off, see Fig. \ref{fig:sharp}(a). This simplified choice was also already discussed in Ref. \cite{cMERA}. Through the condition $d \alpha(k)/d k=2 g(k)\alpha(k)/k$ (proved below), we obtain a simplified function $\alpha(k)$, denoted $\tilde{\alpha}(k)$, 
\begin{equation}  \label{eqApp:alpha_tilde}
  \tilde{\alpha}(k) \equiv \left\{ \begin{array}{cl}
  |k|,&  ~~~~|k| \leq \Lambda,\\
  \Lambda,& ~~~~ |k| >\Lambda,\\
  \end{array}\right.
\end{equation} 
represented in Fig. \ref{fig:sharp}(b), which characterizes the simplified cMERA state $\ket{\tilde{\Psi}^{\cM}}$,
\begin{equation} \label{eqApp:cMERA_tilde}
\left(\sqrt{\frac{\tilde{\alpha}(k)}{2}} \phi(k)+\frac{i}{\sqrt{2\tilde{\alpha}(k)}} \pi(k) \right)\ket{\tilde{\Psi}^{\cM}} =0,~~ \forall k.
\end{equation}

The above simplified $\tilde{g}(k)$ leads to an entangler $\tilde{K}$ that is not quasi-local in real space \cite{cMERA}, as can be readily checked by performing a Fourier transform. That is, the simplified cMERA $\ket{\tilde{\Psi}^{\cM}}$ fails to satisfy a defining feature of a proper cMERA $\ket{\Psi^{\cM}}$, namely that the generator $L+K$ of the entangling evolution be quasi-local. Still, the entangler $\tilde{K}$ leads to a simplified flow in the space of constraints (to be discussed below) that already contains the essential ingredients of the flow that appears when using a quasi-local $K$. 

\subsection{Relativistic and non-relativistic scale transformations}

With $\tilde{g}(k)$, the operator $L+\tilde{K}$ acts as the generator $D$ of CFT dilations  (named relativistic scaling operator $L'$ in Ref. \cite{cMERA}),
\begin{equation}
D \equiv \frac 1 2 \int dk \Big[  \pi(-k) (k \partial_k +1) \phi(k)+ h.c. \Big],
\end{equation}
for $|k| < \Lambda$ and as the non-relativistic scaling operator $L$ in Eq. \ref{eqApp:L} for $|k|>\Lambda$. 

Recall that the CFT dilation generator $D$, which generates scale transformations on the CFT, acts on the field operators as
\begin{eqnarray}
-i[D,\phi(k)] &=& -\left(k\partial_k + 1 \right) \phi(k),\\
-i[D,\pi(k)] &=& -\left(k\partial_k + 0 \right) \pi(k). 
\end{eqnarray}
Therefore, an evolution by $D$ for a small $\epsilon > 0$ transforms the operators $\phi(k)$ and $\pi(k)$ as
\begin{equation} \label{eqApp:x}
\phi(k) \rightarrow e^{-\epsilon}\phi(e^{-\epsilon}k),~~~
\pi(k) \rightarrow \pi(e^{-\epsilon}k),
\end{equation}
It follows that $D$ has the CFT ground state $\ket{\Psi}$ 
\begin{equation} \label{eqApp:CFT}
\left(\sqrt{\frac{|k|}{2}} \phi(k)+\frac{i}{\sqrt{2|k|}} \pi(k) \right)\ket{\Psi} =0,~~ \forall k,
\end{equation}
as a fixed-point, $D\ket{\Psi} \propto \ket{\Psi}$. 

In turn, the non-relativistic scaling operator $L$, which generates scale transformations in a non-relativistic theory, acts on the fields as
\begin{eqnarray}
-i[L,\phi(k)] &=& -\left(k\partial_k + \frac{1}{2} \right) \phi(k),\\
-i[L,\pi(k)] &=& -\left(k\partial_k + \frac{1}{2} \right) \pi(k), 
\end{eqnarray}
or
\begin{equation} \label{eqApp:y}
\phi(k) \rightarrow e^{-\frac{\epsilon}{2}}\phi(e^{-\epsilon}k), ~~~
\pi(k) \rightarrow e^{-\frac{\epsilon}{2}} \pi(e^{-\epsilon}k)
\end{equation}
from where it follows that it has the product state $\ket{\Lambda}$ 
\begin{equation} \label{eqApp:prod}
\left(\sqrt{\frac{\Lambda}{2}} \phi(k)+\frac{i}{\sqrt{2\Lambda}} \pi(k) \right)\ket{\Lambda} =0,~~ \forall k,
\end{equation}
as a fixed-point, $L\ket{\Lambda} \propto \ket{\Lambda}$.

\subsection{Linear constraints and their flow under $L+\tilde{K}$}

Consider now a constraint of the form
\begin{equation} \label{eqApp:constraint1}
\left(\sqrt{\frac{\beta}{2}} \phi(k)+\frac{i}{\sqrt{2\beta}} \pi(k) \right)\ket{\Phi} =0,
\end{equation}
where $\beta>0$ is some real value. We can characterize this linear constraint on the state $\ket{\Phi}$ by a pair $(k,\beta)$. Under a small evolution of the operators $\phi(k)$ and $\pi(k)$ by $L+\tilde{K}$, Eqs. \ref{eqApp:x}-\ref{eqApp:y}, this constraint is transformed into a new constraint
\begin{equation} \label{eqApp:constraint2}
e^{-\frac{\epsilon}{2}}\left(\sqrt{\frac{\beta}{2}} e^{-\frac{\epsilon}{2}}\phi(e^{-\epsilon}k)+\frac{i}{\sqrt{2\beta}} e^{\frac{\epsilon}{2}}\pi(e^{-\epsilon}k) \right)\ket{\Phi} =0,
\end{equation}
if $|k|\leq \Lambda$, and
\begin{equation} \label{eqApp:constraint3}
e^{-\frac{\epsilon}{2}}\left(\sqrt{\frac{\beta}{2}}\phi(e^{-\epsilon}k)+\frac{i}{\sqrt{2\beta}}\pi(e^{-\epsilon}k) \right)\ket{\Phi} =0,
\end{equation}
if $|k| > \Lambda$. Therefore the new constraint 
\begin{equation} \label{eqApp:constraint4}
\left(\sqrt{\frac{\beta'}{2}} \phi(k')+\frac{i}{\sqrt{2\beta'}} \pi(k') \right)\ket{\Phi} =0,
\end{equation}
is characterized by the pair $(k',\beta')$ with $k'\equiv ke^{-\epsilon}$ and
\begin{equation}
\beta' = \left\{
\begin{array}{rc}
e^{-\epsilon}\beta  & \mbox{for} ~ |k| \leq \Lambda, \\
\beta & \mbox{for} ~ |k| > \Lambda.
\end{array}
\right.
\end{equation}
Fig. \ref{fig:sharp}(c) represents the flow generated by $L+\tilde{K}$ in the space of constraints $(k,\beta)$. Notice the ground state $\ket{\Psi}$ in Eq. \ref{eqApp:CFT}, the product state $\ket{\Lambda}$ in Eq. \ref{eqApp:prod}, and the simplified cMERA state $\ket{\tilde{\Psi}^{\cM}}$ in Eq. \ref{eqApp:cMERA_tilde} are all characterized by a complete family of linear constraints $(k,\beta(k))$ for all $k \in \mathbb{R}$, to which we can refer by just specifying the function $\beta(k)$. Specifically, they correspond to $\beta(k) = |k|$, $\beta(k) = \Lambda$, and $\beta(k) = \tilde{\alpha}(k)$, respectively. In particular, it can be seen that the curve $\tilde{\alpha}(k)$ for the simplified cMERA $\ket{\tilde{\Psi}^{\cM}}$ is a fixed-point of this flow.

\begin{figure}[t]
\centering
\includegraphics[width=8.5cm]{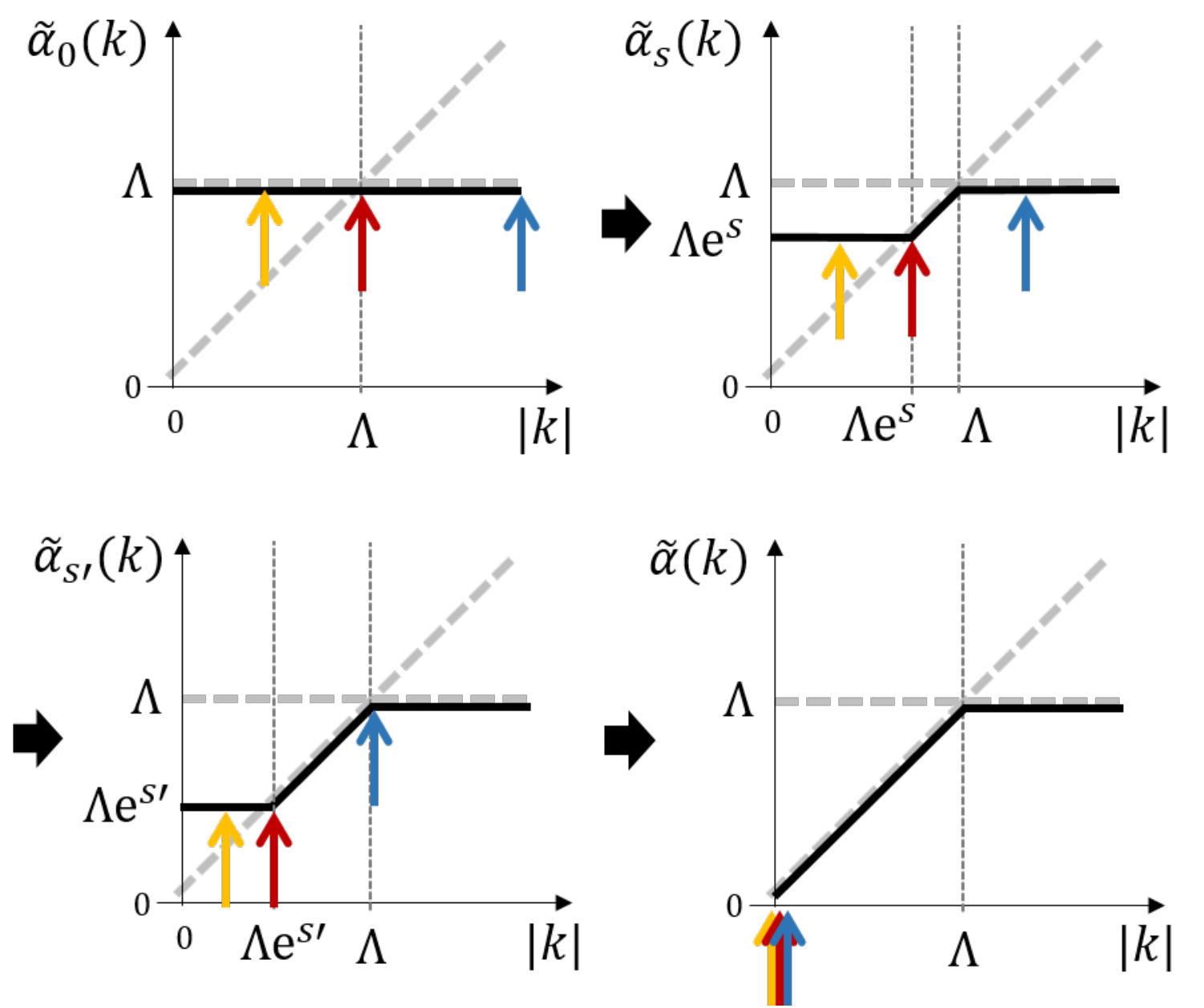}
\caption{
Evolution of a complete set of constraints under a scale transformation in scale generated by the simplified $L+\tilde{K}$. Starting from the product state $\ket{\Lambda}$ in Eq. \ref{eqApp:prod}, the sequence represents the constraints $\tilde{\alpha}_{s_{IR}}(k)$ defining $\ket{\tilde{\Psi}^{\cM}_{s_{IR}}}$ for four different values of $s_{IR} \leq 0$, namely $s_{IR} = 0, s, s', \infty$ , with $0>s>s'> \infty$. Each coloured arrow corresponds to a constraint, which evolves as a function of $s_{IR}$. The function $\tilde{\alpha}(k)$ appears as the fixed-point of the evolution.
}
\label{fig:sharpEvo}
\end{figure}

\subsection{Evolution in scale}

We can now put all the pieces together and study how the characterization of the state
\begin{equation}
\ket{\tilde{\Psi}^{\Lambda}_{s_{IR}}} \equiv  e^{-i(L+\tilde{K})(0-s_{IR})} \ket{\Lambda}
\end{equation}
in terms of a function $\beta(k) = \tilde{\alpha}_{s_{IR}}(k)$ depends on the infrared scale $s_{IR} \leq 0$, which measures the length of the evolution. This is illustrated in Fig. \ref{fig:sharpEvo}. For $s_{IR}=0$ we have the product state $\ket{\tilde{\Psi}^{\cM}_0} = \ket{\Lambda}$, characterized by a flat $\tilde{\alpha}_{0}(k) \equiv \Lambda$. For finite $s_{IR} < 0$, the state $\ket{\tilde{\Psi}^{\cM}_{s_{IR}}}$ is characterized by a function $\tilde{\alpha}_{s_{IR}}(k)$ with two plateaus, one at $\Lambda e^{s_{IR}}$ for small $|k|$ and another at $\Lambda$ for large $|k|$, connected by a segment of linear growth:
\begin{equation}  
  \tilde{\alpha}_{s_{IR}}(k) \equiv \left\{ \begin{array}{cl}
  \Lambda e^{s_{IR}}& ~~~~ \mbox{for}~ |k| \leq \Lambda e^{s_{IR}}, \\
  |k|&  ~~~~\mbox{for}~|k| \in [ \Lambda e^{s_{IR}}, \Lambda],\\
  \Lambda& ~~~~ \mbox{for}~|k| >\Lambda.\\
  \end{array}\right.
\end{equation} 
Then, in the limit $s_{IR} \rightarrow - \infty$ we recover the $\tilde{\alpha}(k)$ of Eq. \ref{eqApp:alpha} characterizing the simplified cMERA state $\ket{\tilde{\Psi}^{\cM}}$. 

We conclude that the simplified cMERA $\ket{\tilde{\Psi}^{\cM}}$ can be indeed equivalently characterized either (i) as the result of applying the unitary evolution $\tilde{U}$ on the product state $\ket{\Lambda}$, $\ket{\tilde{\Psi}^{\cM}} =\tilde{U} \ket{\Lambda}$, where
\begin{equation}
\tilde{U} \equiv \lim_{s_{IR}\rightarrow -\infty} e^{-i(L+\tilde{K})(0-s_{IR})},
\end{equation}
or (ii) in terms of the complete set of linear constraints $(k,\tilde{\alpha}(k))$ for all $k\in\mathbb{R}$, Eqs. \ref{eqApp:alpha_tilde}-\ref{eqApp:cMERA_tilde}.

The connection between the two characterizations can be simply summarized by the observation that $\tilde{U}$ is a symplectic transformation that acts on the fields as
\begin{eqnarray}
\tilde{U}\phi(k)\tilde{U}^{\dagger} &=& \sqrt{\frac{\tilde{\alpha}(k)}{\Lambda}}\phi(k), \label{eqApp:symp1}\\
\tilde{U}\pi(k)\tilde{U}^{\dagger} &=&  \sqrt{\frac{\Lambda}{\tilde{\alpha}(k)}}\pi(k), \label{eqApp:symp2}
\end{eqnarray}
so that it transforms the complete set of constraints  $\beta(k)= \Lambda$ of the product state $\ket{\Lambda}$ into the complete set of constraints $\beta(k) = \tilde{\alpha}(k)$ of the simplified cMERA state $\ket{\tilde{\Psi}^{\cM}}$, namely
\begin{eqnarray}
&&\tilde{U}\left(\sqrt{\frac{\Lambda}{2}} \phi(k)+\frac{i}{\sqrt{2\Lambda}} \pi(k) \right)
\tilde{U}^{\dagger} = \\
&&~~~~~~~~\left(\sqrt{\frac{\tilde{\alpha}(k)}{2}} \phi(k)+\frac{i}{\sqrt{2\tilde{\alpha}(k)}} \pi(k) \right)
\end{eqnarray}

Analogous statements apply also for the choice of $g(k)$ leading to a quasi-local $K$, such as the one used in the main text, as discussed next.
 
\subsection{Connecting $g(k)$ and $\alpha(k)$}

Let us now consider a more general $g(k)$ still satisfying that it approaches $1/2$ and $0$ at very small and very large momenta,
\begin{equation} \label{eqApp:g}
  g(k) \sim \left\{ \begin{array}{cl}
  1/2,&  ~~~~|k|\ll \Lambda,\\
  0,& ~~~~ |k| \gg \Lambda,
  \end{array}\right.
\end{equation} 
so that $L+K$ approaches the CFT dilation generator $D$ and the non-relativistic scaling operator $L$ in those two limits. The fixed-point of the unitary evolution generated by this $L+K$ is then a cMERA that can be alternatively characterized by a complete set of linear constraints $\alpha(k)$, where one expects $\alpha(k)$ to approach the behaviour of the constraints for the CFT ground state $\ket{\Psi}$ and that of the constraints for the product state $\ket{\Lambda}$ at very small and very large momenta,
\begin{equation} \label{eqApp:alpha}
  \alpha(k) \sim \left\{ \begin{array}{cl}
  |k|,&  ~~~~|k|\ll \Lambda,\\
  \Lambda,& ~~~~ |k| \gg \Lambda.
  \end{array}\right.
\end{equation}  

Next we show that given a function $g(k)$ for the entangler $K$, the function $\alpha(k)$ characterizing a complete set of constraints for the resulting cMERA $\ket{\Psi^{\cM}}$, namely
\begin{equation} \label{eqApp:cMERA2}
a^{\cM}(k) \ket{\Psi^{\cM}} =0,~~ \forall k,
\end{equation}
with
\begin{equation}\label{eqApp:cMERA}
a^{\cM}(k) \equiv \sqrt{\frac{\alpha(k)}{2}} \phi(k)+\frac{i}{\sqrt{2\alpha(k)}} \pi(k),
\end{equation}
is the solution to the differential equation
\begin{equation} \label{eqApp:galpha}
 \frac{d\alpha(k)}{dk} =\frac{ 2g(k)\alpha(k)}{k}.
\end{equation} 

This is shown by studying how the annihilation operator $a^{\cM}(k)$ evolves under $D^{\cM} = L+K$. Explicitly, since 
\begin{eqnarray}
-i[D^{\cM},\phi(k)] &=& -\left(k\partial_k + \frac{1}{2} + g(k) \right) \phi(k),\\
-i[D^{\cM},\pi(k)] &=& -\left(k\partial_k + \frac{1}{2} - g(k)\right) \pi(k),
\end{eqnarray}
we then have that
\begin{eqnarray}
&&-i\left[ D^{\cM}, a^{\cM}(k)\right]\\
&=& -i\left[D^{\cM},\sqrt{\frac{\alpha(k)}{2}}\phi(k) + i  \sqrt{\frac{1}{2\alpha(k)}}\pi(k)\right]\\
&=& -\sqrt{\frac{\alpha(k)}{2}}\left(k\partial_k +\frac{d}{2} + g(k)\right)\phi(k) \\
&& -i \sqrt{\frac{1}{2\alpha(k)}}\left(k\partial_k +\frac{d}{2} - g(k)\right)\pi(k) \\
&=& -\left(k\partial_k +\frac{d}{2}\right)\left[ \sqrt{\frac{\alpha(k)}{2}}\phi(k) + i  \sqrt{\frac{1}{2\alpha(k)}}\pi(k)\right]\\
&+&\left(\frac{k}{2\alpha(k)}\frac{d\alpha(k)}{dk} - g(k)\right)\left[ \sqrt{\frac{\alpha(k)}{2}}\phi(k) - i  \sqrt{\frac{1}{2\alpha(k)}}\pi(k) \right] \nonumber\\
&=& -\left(k\partial_k +\frac{d}{2}\right)a^{\cM}(k)
+\left(\frac{k}{2\alpha(k)}\frac{d\alpha(k)}{dk} - g(k)\right)a^{\cM}(-k)^{\dagger}. \nonumber
\end{eqnarray}

This expression reduces to 
\begin{equation} \label{eqApp:change}
 -i\left[ D^{\cM}, a^{\cM}(k)\right] = -\left(k\partial_k +\frac{d}{2}\right)a^{\cM}(k)
\end{equation}
if and only if Eq. \ref{eqApp:galpha} is satisfied. Eq. \ref{eqApp:change} then tells us that under $D^{\cM}$, the linear constraints of Eq. \ref{eqApp:cMERA} are modified by a linear combination of the same constraints. In other words, the linear constraints before and after the transformation by $D^{\cM}$ are collectively equivalent. Therefore they uniquely characterizes the same Gaussian state $\ket{\Psi^{\cM}}$.
 
\subsection{Symplectic transformation $V$}

Let us now examine the symplectic transformation $V$ that maps the target ground state $\ket{\Psi}$ of the 1+1 free boson CFT into the cMERA $\ket{\Psi^{\cM}}$, $\ket{\Psi^{\cM}} = V \ket{\Psi}$. It is given by
\begin{eqnarray}
V\phi(k)V^{\dagger} &=& \sqrt{\frac{\alpha(k)}{|k|}}\phi(k) \equiv \phi^{\cM}(k), \label{eqApp:symp1}\\
V\pi(k)V^{\dagger} &=&  \sqrt{\frac{|k|}{\alpha(k)}}\pi(k) \equiv \pi^{\cM}(k), \label{eqApp:symp2}
\end{eqnarray}
which implies that, indeed, $Va(k)V^{\dagger} = a^{\cM}(k)$, so that it maps the linear constraints of $\ket{\Psi}$ into the linear constraints of $\ket{\Psi^{\cM}}$. 

We do so for the specific choice 
\begin{equation}
g(k) = \frac{1}{2}\exp\left(-\frac{1}{\sigma} \left(\frac{k}{\Lambda}\right)^2 \right),
\end{equation}
which is equivalent to that proposed in Ref. \cite{cMERA}, except for the presence of a factor $\sigma\approx 1.78107$, namely the exponential of Euler's constant $\gamma \approx 0.57722 $. Notice that $g(k)$ tends to $1/2$ and $0$ for small and large $|k|$, as in Eq. \ref{eqApp:g}. The corresponding $\alpha(k)$ satisfying Eq. \ref{eqApp:galpha} is
\begin{equation}
\alpha(k) =\Lambda \exp\left(  \frac 1 2 \textrm{Ei}\Big(-\frac{1}{\sigma} (\frac{k}{\Lambda})^2\Big)         \right),
\end{equation}
where Ei is the special function known as exponential integral,
\begin{equation}
\mbox{Ei} (y) \equiv - \int_{-y}^{\infty} \frac{e^{-t}}{t}dt,
\end{equation}
which accepts the following convergent series
\begin{equation}
\mbox{Ei} (y) = \gamma + \ln|y| + \sum_{k=1}^{\infty} \frac{y^k}{k~k!}.
\end{equation}
From these expressions in can be easily seen that $\alpha(k)$ tends to $|k|$ and $\Lambda$ for small and large $|k|$, as in Eq. \ref{eqApp:alpha}. [We note that $\sigma$ is required in the above expressions in order for $\alpha(k)$ to grow as $\alpha(k) = |k| + O(k^2)$ in the limit of small $k$, as needed if $\ket{\Psi^{\cM}}$ is to approximate $\ket{\Psi}$ at large distances.]

\subsection{Quasi-local scaling operators}

Next we show that the absolute value of the (generalized) functions $\mu_{\phi}(x)$ and $\mu_{\pi}(x)$ for the smeared fields $\phi^{\cM}(x)$ and $\pi^{\cM}(x)$, 
\begin{eqnarray} \label{eqApp:mu_phi}
\phi^{\cM}(x) = V\phi(x)V^{\dagger} = \int dy~  \mu_\phi(x-y) \phi(y),\\
\pi^{\cM}(x) = V\pi(x)V^{\dagger} = \int dy~  \mu_\pi(x-y) \pi(y), \label{eqApp:mu_pi}
\end{eqnarray}
can be upper bounded by $O(\exp\left(-|\Lambda x|\right)$ for $|\Lambda x|\gg 1$, so that we can think of  $\phi^{\cM}(x)$ and $\pi^{\cM}(x)$ as quasi-local fields with a characteristic length scale upper bounded by $1/\Lambda$.

The Fourier transforms of $\sqrt{\frac{\alpha(k)}{|k|}}$ and $\sqrt{\frac{|k|}{\alpha(k)}}$, see Eqs. \ref{eqApp:symp1}-\ref{eqApp:symp2} are distributions and, as such, cannot be reliably obtained through a numerical Fourier transform. [That the profile functions $\mu_{\phi}(x)$ and $\mu_{\pi}(x)$ are actually distributions should not come as a surprise. In the original CFT, with sharp field operators $\phi(x)$ and $\pi(x)$, they are already distribitions, namely delta functions $\delta(x)$]. In order to analyse these distributions, we decompose them as sum of two pieces
\begin{eqnarray}
\mu_\phi(x) &=& \mu_\phi^{(1)}(x)+\mu_\phi^{(2)}(x),\label{eqApp:muphi12}\\
\mu_\pi(x) &=& \mu_\pi^{(1)}(x)+\mu_\pi^{(2)}(x),\label{eqApp:mupi12}
\end{eqnarray}
where the first term is conveniently chosen such that (i) it is a distribution with an analytic Fourier transform and (ii) the remaining term is sufficiently regular that it can be reliably obtained by a numerical Fourier transform.

Since $\sqrt{\frac{\alpha(k)}{|k|}}\sim|k/\Lambda|^{-\frac 1 2}$ as $|k|\to \infty$, we choose $\mu_{\phi}^{(1)}(k) \equiv (1+k^2/\Lambda^2)^{-\frac 1 4}$, which has the same asymptotic behavior for large $|k|$ (to leading order in $|k|$). This function has an analytical distributional Fourier transform,
\begin{equation}
\mu_\phi^{(1)}(x)\equiv\mathcal F\left((1+k^2/\Lambda^2)^{-\frac 1 4}\right)(x)=\frac{2^{3/4} K_{\frac{1}{4}}(|\Lambda x|)}{\Gamma \left(\frac{1}{4}\right) \left| \Lambda x\right|^{1/4} },
\end{equation}
where $K_n$ is the modified Bessel function of the second kind, and $\Gamma$ is the Euler gamma function. It can be shown that it has asymptotic behavior
\begin{equation}
\mu_\phi^{(1)}(x)\sim
\left\{ \begin{array}{ll}
|\Lambda x|^{-1/2} ,&|\Lambda x|\ll 1;\\
~\\
\frac{{2}^{1/4} \sqrt{\pi } e^{-\left| \Lambda x\right| }}{\Gamma \left(\frac{1}{4}\right) \left| \Lambda x\right| ^{3/4}}, & |\Lambda x|\gg 1.
\end{array} \right.
\end{equation}
Arguing similarly about $\sqrt{\frac{|k|}{\alpha(k)}}$, we choose $\mu_{\pi}^{(1)}(k) \equiv (1+k^2/\Lambda^2)^{\frac 1 4}$, which has analytical distributional Fourier transform,
\begin{equation}
\mu_\pi^{(1)}(x)\equiv\mathcal F\left((1+k^2/\Lambda^2)^{\frac 1 4}\right)(x)=\frac{ {2}^{ 5/4} K_{\frac{3}{4}}(\left| \Lambda x\right| )}{\Gamma \left(-\frac{1}{4}\right) \left| \Lambda x\right| ^{3/4}}.
\end{equation}
Its asymptotic behavior is
\begin{equation}
\mu_\pi^{(1)}(x)\sim
\left\{ \begin{array}{ll}
-\frac 1 2 |\Lambda x|^{-3/2} ,&|\Lambda x|\ll 1;\\
~\\ 
\frac{2^{3/4} \sqrt{\pi } e^{-\left| \Lambda x\right| }}{\Gamma \left(-\frac{1}{4}\right) \left| \Lambda x\right| ^{5/4}},& |\Lambda x|\gg 1.
\end{array} \right.
\end{equation}
$\mu_\pi^{(1)}(x)$ should be understood as a distribution. Its action on a test function $f(x)$ is defined as the Hadamard finite-part integral \cite{Hadamard}:
\begin{equation}
\begin{aligned}
&\langle   \mu_\pi^{(1)}, f\rangle \\
&\equiv \lim_{\epsilon\to 0^+}\left( \int_{\mathbb R\setminus(-\epsilon,\epsilon)}  \mu_\pi^{(1)}(x)f(x)dx  +2\Lambda^{-\frac 3 2}\epsilon^{-\frac  1 2} f(0)  \right).
\end{aligned}
\end{equation}
It is clear that $\mu_\phi^{(1)}(x)$ and $\mu_\pi^{(1)}(x)$ both decay exponentially for large $|x|$. Since $\mu_\phi^{(2)}(x)$ and $\mu_{\pi}^{(2)}(x)$ are non-singular, the singular behaviors of $\mu_\phi(x)$ and $\mu_\pi(x)$ for $|\Lambda x|\ll 1$ are the same as $\mu_\phi^{(1)}(x)$ and $\mu_\pi^{(1)}(x)$.

\begin{figure}[t]
\centering
\includegraphics[width=8.0cm]{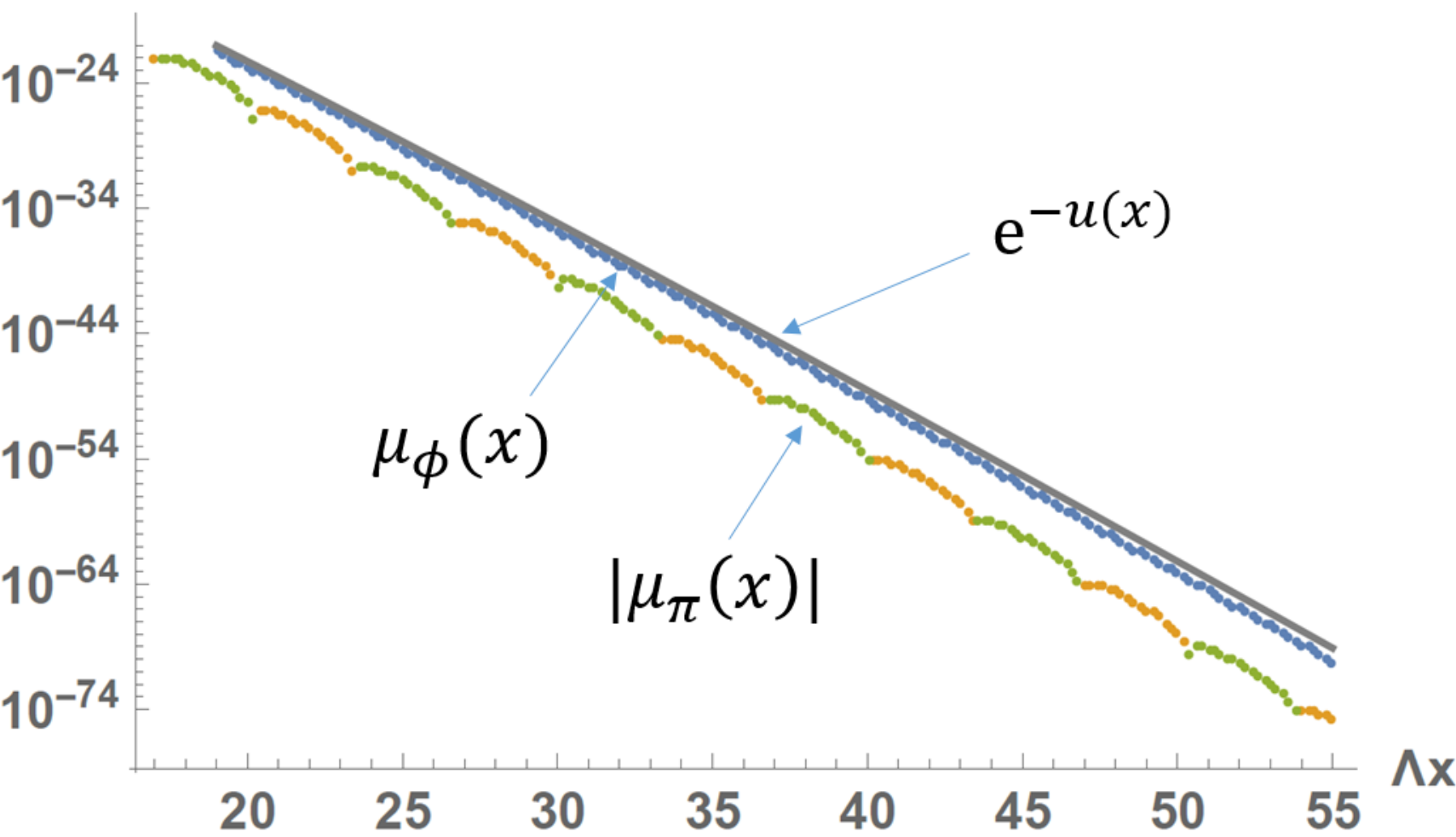}
\caption{Generalized functions $\mu_{\phi}(x)$ and $\mu_{\pi}(x)$ for large $\Lambda x \gg 1$, together with function $const. \times e^{-u(x)}$ for $u(x) =\Lambda x  \sqrt{\sigma\log(\Lambda x)}$. Both $\mu_{\phi}(x)$ (blue) and $|\mu_{\pi}(x)|$ (in yellow when $\mu_{\pi}(x) >0$ and green when $\mu_{\pi}(x) < 0$) where obtained numerically by adding the contributions (see Eqs. \ref{eqApp:muphi12}-\ref{eqApp:mupi12}) of an analytical Fourier transform and a numerical Fourier transform. The results were obtained using arbitrary precision arithmetic.
}
\label{fig:smeared_fields}
\end{figure}

After the subtraction, we can readily apply a numerical Fourier transform to $\mu_{\phi}^{(2)}(k) \equiv \sqrt{\frac{\alpha(k)}{|k|}}-(1+k^2/\Lambda^2)^{-\frac 1 4}$ and $\mu_{\pi}^{(2)}(k) \equiv \sqrt{\frac{|k|}{\alpha(k)}}-(1+k^2/\Lambda^2)^{\frac 1 4}$. By summing the analytical part and numerical part, we can see that $\mu_\phi(x)$ and $\mu_\pi(x)$ decay roughly esxponentially for large $x$, and $\mu_\pi(x)$ also oscillates, as shown in Fig. \ref{fig:smeared_fields}.

%

In fact, it can be proven that 
\begin{equation} \label{eqApp:Asymptotic}
\mu_\phi(x), \mu_\pi(x)= 
\text{exp} \left\{ -\Lambda x \sqrt{\sigma \log (\Lambda x)}+o(\Lambda x\sqrt{\log (\Lambda x)})   \right\},
\end{equation}
as   $x\to \infty$. The rigorous proof is rather lengthy, so here we only outline it for the asymptotic behavior of $\mu_\phi(x)$.
That $\mu_\phi(x)$ is the Fourier transform of $\sqrt{\frac{\alpha(k)}{|k|}}$ is equivalent to the following integral equation,
\begin{equation}
(x\partial_x+\frac 1 2)\mu_\phi(x)+\frac{1}{\sqrt{2\pi}}\int g(x-y)\mu_\phi(y)dy=0,
\label{integral_eq}
\end{equation}
where $g(x)=\sqrt{\frac{\sigma \Lambda^2}{8}}e^{-\frac 1 4  \sigma \Lambda^2 x^2}$ is the Fourier transform of $g(k)$. (This follows from requiring that $-i[D^{\cM}, \phi^{\cM}(x)] = (x\partial_x + 0)\phi^{\cM}(x)$ at $x=0$). We define a new variable $u(x)$ such that,
\begin{equation}
\mu_\phi(x)=e^{-u(x)}.
\end{equation}
We seek the solution to this integral equation among functions satisfying a few conditions:
\begin{eqnarray}
\begin{aligned} 
&u'(x)\to \infty,  ~~ \text{log}(u'(x))=o(\text{log}(\Lambda x)),  \text{   as   }x\to \infty;\\
&|u'(x)-u'(y)|=o(\sqrt{\text{log}(\Lambda x)})+o(\Lambda |x-y|), \\
&~~~~~~~~~~~~~~~~~~~~~~~~~~~~~~~~~~~~~~~~~~~~~ \text{   as   }x,y \to \infty.
\label{condition} 
\end{aligned}
\end{eqnarray}
Then the convolution term in Eq. \ref{integral_eq} can be approximated as 
\begin{eqnarray}
&&\int g(x-y)\mu_\phi(y)dy \nonumber \\
&&~~~~ =\sqrt{\frac \pi 2}\mu_\phi(x)e^{\frac {1}{\sigma \Lambda^2}\big(u'(x)+o(\sqrt{\log \Lambda x})\big)^2(1+o(1))}.
\end{eqnarray}
One can show that there exists solution to Eq. \ref{integral_eq} satisfying conditions Eq. \ref{condition}, and it reads
\begin{equation}
u(x)=\Lambda x \sqrt{\sigma \log \Lambda x}+o(\Lambda x\sqrt{ \log \Lambda x}).
\end{equation}
Then we prove Eq. \ref{eqApp:Asymptotic} for $\mu_\phi(x)$. The proof for $\mu_{\pi}(x)$ is obtained similarly, starting from $-i[D^{\cM}, \pi^{\cM}(x)] = (x\partial_x + 1)\pi^{\cM}(z)$ at $x=0$.


\section{APPENDIX B: Radial and N-S quantizations}

In this paper we have analysed the spacetime symmetries of the cMERA state $\ket{\Psi^{\cM}}$ for the ground state $\ket{\Psi}$ of a 1+1 dimensional CFT where space corresponds to the real line $x\in \mathbb{R}$. We have done so in terms of a subset of generators of global conformal transformations on the real line (namely the Hamiltonian $H$, the momentum operator $P$, the dilation and boost generators $D$ and $B$, which can be completed with the generators of special conformal transformations $K_1$ and $K_2$). 

The spacetime symmetries of a 1+1 CFT are usually analysed using instead radial quantization, where space corresponds to the circle. The Virasoro generators $L_n$ and $\bar{L}_n$ can then be mapped onto generators $L'_n$ and $\bar{L}'_n$ on the real line, producing the so-called N-S quantization (North pole - South pole quantization), as described e.g. in S. Rychkov's notes \textit{EPFL Lectures on Conformal Field Theory in $D>= 3$ Dimensions}, arXiv:1601.05000, for the global conformal group in higher dimensions. The goal of this appendix, which contains no original research, is to briefly review the connection between radial quantization and N-S quantization, following a detailed explanation kindly offered to the authors by John Cardy, and to clarify their relation to the generators $H,P,B,D$ used in the main text.

\subsection{Complex coordinates and stress-energy tensor}
We can parameterize the Euclidean plane by the complex coordinate $z \equiv z^0+iz^1$, and use $z$ and $\bar{z}$ as new coordinates,
\begin{equation}
\begin{array}{ll}
z = z^0 + iz^1~~~~~ & z^0 = \frac{1}{2}\left(z + \bar{z}\right)\\
&\\
\bar{z} = z^0 - iz^1~~~~~ & z^1 = \frac{1}{2i}\left(z - \bar{z}\right)\\
&\\
\partial_z = \frac{1}{2} \left(\partial_0 - i\partial_1 \right) ~~~~~& \partial_0 = \partial_z + \partial_{\bar{z}} \\
& \\
\partial_{\bar{z}} = \frac{1}{2} \left(\partial_0 + i\partial_1 \right) ~~~~~& \partial_1 = i\left(\partial_z - \partial_{\bar{z}}\right) 
\end{array}
\end{equation}

The stress-energy tensor $T_{\mu\nu}(z^0,z^1)$ of a CFT is symmetric ($T_{10} = T_{01}$) and traceless ($T_{11}=-T_{00}$). In complex coordinates we obtain components
\begin{eqnarray}
T_{zz} &=& \frac{1}{4}\left(T_{00} -i(T_{10} + T_{01}) - T_{11}\right) \\
&=&  \frac{1}{2} \left(T_{00}  - i T_{01} \right)\\
T_{\bar{z}\bar{z}} &=& \left(T_{00} +i(T_{10} + T_{01}) - T_{11}\right)\\
&=&  \frac{1}{2} \left(T_{00}  + i T_{01} \right) 
\end{eqnarray}
whereas $T_{z\bar{z}} = T_{\bar{z}z} = (T_{00}+T_{11})/4 =0$. In addition, from the conservation law $g^{\alpha \mu}\partial_{\alpha}T_{\mu\nu} = 0$ it follows that $T_{zz}$ is a holomorphic function ($\partial_{\bar{z}}T_{zz} = 0$) whereas $T_{\bar{z}\bar{z}}$ is antiholomorphic ($\partial_z T_{\bar{z}\bar{z}}=0$). Finally, the \textit{renormalized} holomorphic and antiholomorphic components of the stress-energy tensor are given by
\begin{eqnarray}
 T(z) \equiv -2\pi T_{zz}(z), ~~~~~\bar{T}(\bar{z}) \equiv -2\pi T_{\bar{z}\bar{z}} (\bar{z})
\end{eqnarray}

\subsection{Generators of conformal coordinate transformation}

Holomorphic (antiholomorphic) conformal coordinate transformations $z \rightarrow f(z)$ (respectively $\bar{z} \rightarrow \bar{f}(\bar{z})$) preserve angles and are generated by infinitesimal transformations 
\begin{eqnarray}
z\rightarrow z + \sum_{n=-\infty}^{\infty}c_n z^{n+1} = (1 - \sum_{n=-\infty}^{\infty} c_n l_n)z \\
\bar{z}\rightarrow \bar{z} + \sum_{n=-\infty}^{\infty}\bar{c}_n \bar{z}^{n+1} = (1 - \sum_{n=-\infty}^{\infty} \bar{c}_n \bar{l}_n)\bar{z},
\end{eqnarray}
where $c_n \in \mathbb{C}$ and the holomorphic and antiholomorphic generators $l_n$ and $\bar{l}_n$ are given by
\begin{eqnarray}
 l_n \equiv -z^{n+1}\partial_z,~~~~~ \bar{l}_n \equiv -\bar{z}^{n+1}\partial_{\bar{z}},
\end{eqnarray}
which close the Witt algebra \cite{CFTbook}
\begin{eqnarray}
\left[l_m,l_n\right] &=& (m-n)l_{m+n},\\
\left[\bar{l}_m,\bar{l}_n\right] &=& (m-n)\bar{l}_{m+n},\\
\left[l_m,\bar{l}_n\right] &=&  0.
\end{eqnarray}

\subsection{Radial quantization}

In radial quantization, the Virasoro generators are defined in terms of the (renormalized) holomorphic and antiholomorphic components $T(z)$ and $\bar{T}(\bar{z})$ of the stress-energy tensor as
\begin{eqnarray}
L_n &\equiv& \frac{1}{2\pi i} \oint_{|z|=1} dz~z^{n+1} T(z),\\
\bar{L}_n &\equiv& \frac{1}{2\pi i} \oint_{|\bar z|=1} d\bar{z} ~ \bar{z}^{n+1} \bar{T}(\bar{z}),
\end{eqnarray}
where the integrals are over the unit circle $|z|=1$ and $|\bar z|=1$ respectively.
They close the Virasoro algebra \cite{CFTbook} 
\begin{eqnarray}
\left[L_m,L_n\right] &=& (m-n)L_{m+n} +\frac{c}{12}m(m^2-1)\delta_{m,-n},~~~~ \label{eqApp:Virasoro1}\\
\left[\bar{L}_m,\bar{L}_n\right] &=&(m-n)\bar{L}_{m+n} +\frac{c}{12}m(m^2-1)\delta_{m,-n},~~~~\\
\left[L_m,\bar{L}_n\right] &=& 0.~~~~\label{eqApp:Virasoro3}
\end{eqnarray}


\subsection{From the circle to the real line}

\begin{figure}[t]
\centering
\includegraphics[width=8.5cm]{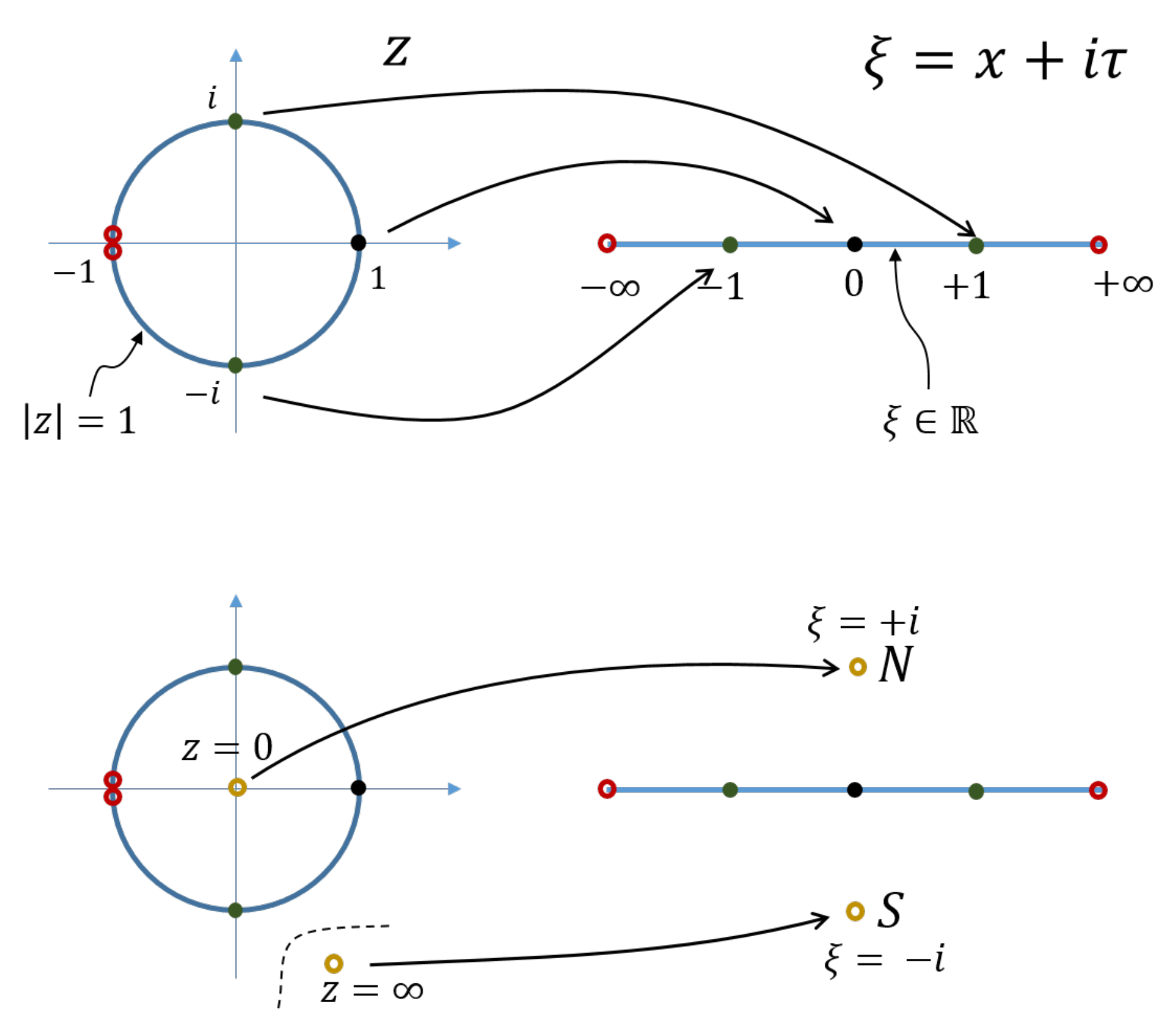}
\caption{
Under the conformal map $z\rightarrow \xi \equiv x+i\tau$, where $z=(1+i\xi)/(1-i\xi)$, the unit circle $|z|=1$ is mapped into the real line $\xi = x$, where $x \in \mathbb{R}$. 
}
\label{fig:zxi}
\end{figure}

We would like to obtain an expression for the generators of the conformal group when the CFT is quantized on the real line. The usual strategy is to map the generators $L_n$ and $\bar{L}_n$ from the unit circle to the real line. For this purpose, consider the conformal map
\begin{equation}
z \rightarrow \xi(z) = x(z) + i\tau(z),~~~~~z = \frac{1+i\xi}{1-i\xi},
\end{equation}
which indeed maps the unit circle $|z|=1$ to the real line $\xi = x \in \mathbb{R}$, as illustrated in Fig. \ref{fig:zxi}. Specifically, the point $z=1$ is mapped to the origin of the real line $x=0$, whereas $\lim_{\theta \rightarrow \pi^{\mp}} e^{i\theta}$ are mapped to to $x=\pm\infty$. Notice also that the origin $z=0$ is mapped to $\xi = +i$, which we will refer to as North pole $N$, while $z=\infty$ is mapped to $\xi = -i$, which we will refer to as South pole $S$. 

Notice that
\begin{equation}
\frac{d z}{d \xi} = \frac{2i}{\left(1-i\xi\right)^2},~~~~ \mbox{or} ~~~~\partial_z =\frac{\left(1-i\xi\right)^2}{2i} \partial_{\xi}
\end{equation}
and therefore under $z \rightarrow \xi$ the generator $l_n$ is mapped into the generator $q_n$ 
\begin{eqnarray}
 q_n &\equiv& - \left(\frac{1+i\xi}{1-i\xi}\right)^{n+1} \frac{(1-i\xi)^2}{2i} \partial_\xi \\
&=&\frac{i}{4}(1+i\xi)^{n+1}(1-i\xi)^{-n+1}\frac{1}{2}(\partial_x -i\partial_\tau) 
\end{eqnarray}
where we used $\partial_{\xi} = \frac{1}{2}(\partial_x -i\partial_\tau)$. Specializing the holomorphic generator $q_n$ to the real line $\xi = x$, and switching from Euclidean time $\tau$ to Lorentzian time $t$, where $\tau=it$, so that $\partial_{\xi} = \frac{1}{2}(\partial_x +\partial_t)$, we finally obtain the right moving generator
\begin{equation}
q_n = \frac{i}{4}(1+ix)^{n+1}(1-ix)^{-n+1}(\partial_x +\partial_t).
\end{equation}
Similarly, from $\bar{z} = (1-i\bar{\xi})/(1+i\bar{\xi})$, we can build antiholomorphic generators, which after switching to Lorentzian time turn into the left moving generators
\begin{equation}
\bar{q}_n = \frac{-i}{4}(1-ix)^{n+1}(1+ix)^{-n+1}\left(\partial_x - \partial_t \right).
\end{equation}

For later reference, we write explicitly the generators of global conformal transformation, namely 
\begin{eqnarray}
q_{0} &=& \frac{i}{4} \left(1+x^2\right) (\partial_x +\partial_t),\\
q_{\pm 1}&=& \frac{i}{4} \left(1\pm ix\right)^2 (\partial_x +\partial_t),\\
\bar{q}_{0} &=& \frac{-i}{4} \left(1+x^2\right) (\partial_x -\partial_t),\\
\bar{q}_{\pm 1}&=& \frac{-i}{4} \left(1 \mp ix\right)^2 (\partial_x -\partial_t).
\end{eqnarray}

On the other hand, under the Moebius transformation $z\rightarrow \xi$, the stress-energy tensor changes simply as \cite{CFTbook}
\begin{equation}
T(z)dz = \frac{T(\xi)d\xi}{dz/d\xi} =  \frac{(1-i\xi)^2}{2i}T(\xi)d\xi.
\end{equation}
Accordingly, the generators of conformal transformations become
\begin{eqnarray}
Q_n &=& \frac{-1}{4\pi} \int dx~\left(\frac{1+ix}{1-ix}\right)^{n+1}(1-ix)^2 T(x)\\
&=& \frac{-1}{4\pi} \int dx~(1+ix)^{n+1}(1-ix)^{-n+1 } T(x).
\end{eqnarray}
where $\int dx \equiv  \int_{\mathbb{R}} d\xi$.
Similar derivations for the antiholomorphic component $\bar{T}(z)$ of the stress-energy lead to
\begin{eqnarray}
\bar{Q}_n = \frac{-1}{4\pi} \int dx~(1-ix)^{n+1}(1+ix)^{-n+1 } \bar{T}(x)
\end{eqnarray}

In particular, for $n=0,\pm 1$ we have the generators of the global conformal transformations, 
\begin{eqnarray}
Q_{0} &=& \frac{-1}{4\pi} \int dx~ (1 + x^2) T(x),\\
Q_{\pm 1} &=& \frac{-1}{4\pi} \int dx~ (1\pm ix)^{2} T(x),\\
\bar{Q}_{0} &=& \frac{-1}{4\pi} \int_{\mathbb{R}} dx~ (1 + x^2) \bar{T}(x),\\
\bar{Q}_{\pm 1} &=& \frac{-1}{4\pi} \int dx~ (1\mp ix)^{2} \bar{T}(x).
\end{eqnarray}

\begin{figure}[t]
\centering
\includegraphics[width=6.0cm]{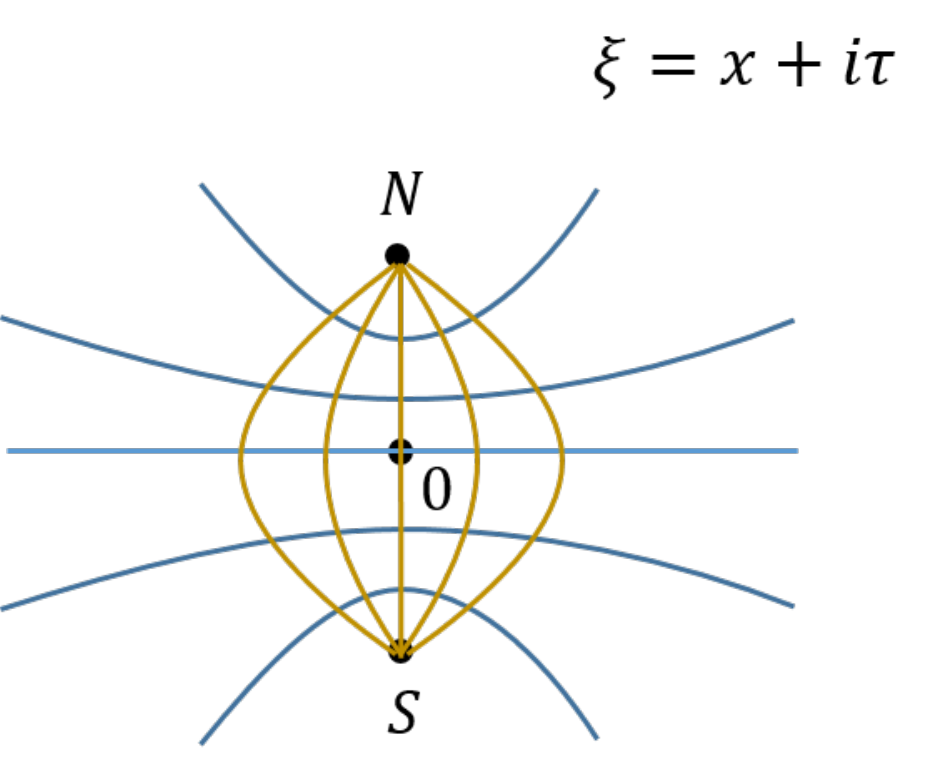}
\caption{
In N-S quantization, the Euclidean plane $\xi=x+i\tau$ is foliated with regards to the action of $L'_0 + \bar{L}_0'$, which in Hilbert space on the real line $\tau=0$ acts as $H+K_1 = -i(1+x^2)\partial_{\tau}$ 
}
\label{fig:NS}
\end{figure}

\subsection{Global conformal transformations directly on the real line}

Let us now consider conformal transformations acting on Minkowski space $\xi = x-t$. We specialize to their action on the $x$ axis, that is for $t=0$. For simplicity, we restrict our attention to global conformal transformations,
\begin{equation}
\begin{array}{ll}
(x,0) \rightarrow (x,t_0) & ~~~~\mbox{time translations} \label{eqApp:coordinategenerators1}\\
(x,0) \rightarrow (x+x_0,0) & ~~~~\mbox{space translations}\\
 (x,0)\rightarrow \gamma (x,-vx) & ~~~~\mbox{boosts} \\
(x,0)\rightarrow (\lambda x, 0) & ~~~~\mbox{dilations}\\
(x,0) \rightarrow (x,ax^2)& ~~~~\mbox{special conformal 1} \\ 
(x,0) \rightarrow (\frac{x}{1-bx},0) &~~~~\mbox{special conformal 2}
\end{array}
\end{equation}
where $t_0,x_0,\gamma, v, a, b$ are real parameters. They are generated, both as a coordinate transformation and as a Hilbert space transformation, by
\begin{equation}
\begin{array}{rll}
i\partial_t, &~ H \equiv \int dx~ h(x)& ~~\mbox{Hamiltonian} \\
-i\partial_x, &~ P \equiv \int dx~ p(x) &~~\mbox{momentum generator}\\
ix\partial_t, &~ B \equiv \int dx~ x~h(x)&~~\mbox{boost generator} \\
-ix\partial_x, &~ D \equiv \int dx~ x~p(x)&~~\mbox{dilation generator}\\
ix^2\partial_t, &~ K_1 \equiv \int dx~ x^2~h(x) &~~\mbox{SC generator 1} \\ 
-ix^2\partial_x, &~ K_2 \equiv \int dx~ x^2~p(x)&~~\mbox{SC generator 1}
\end{array}
\end{equation}
where we have expressed the generators in terms of the Hamiltonian and momentum densities $h(x)$ and $p(x)$, 
\begin{eqnarray}
h(z) &=& \frac{-1}{2\pi} \left( T(z)+\bar{T}(\bar{z})\right),\\
p(z) &=& \frac{-1}{2\pi} \left(T(z)-\bar{T}(\bar{z}) \right).
\end{eqnarray}
These generators fulfil the expected commutation relations, including
\begin{eqnarray}
&&[H,P] = 0,~~~~~[B,D] = 0,\\
&&[B,H]= iP,~~~~[B,P] =iH,\\
&&[D,H]= iH,~~~[D,P] =iP.\\
\end{eqnarray}

The above two sets of global conformal generators on the real line can, of course, be written in terms of each other. Explicitly, we find
\begin{eqnarray}
&&q_0+\bar{q}_0 = \frac{1+x^2}{2}i \partial_t\\
&&q_0-\bar{q}_0 = \frac{1+x^2}{2}i \partial_x\\
&&\left(q_1+q_{-1}\right) +  \left(\bar{q}_1+\bar{q}_{-1}\right)= (1-x^2)i\partial_t\\
&&\left(q_1+q_{-1}\right) -  \left(\bar{q}_1+\bar{q}_{-1}\right)= (1-x^2)i\partial_x\\
&&\left(q_1-q_{-1}\right) +  \left(\bar{q}_1-\bar{q}_{-1}\right)= -2x\partial_t\\
&&\left(q_1-q_{-1}\right) -  \left(\bar{q}_1-\bar{q}_{-1}\right)= -2x\partial_x
\end{eqnarray}
and, correspondingly,
\begin{eqnarray}
&&Q_0 + \bar{Q}_0 = \frac{H+K_1}{2} \\
&&Q_0 - \bar{Q}_0 = -\frac{P+K_2}{2} \\
&&\left(Q_1+Q_{-1}\right) +  \left(\bar{Q}_1+\bar{Q}_{-1}\right)= H-K_1\\
&&\left(Q_1+Q_{-1}\right) -  \left(\bar{Q}_1+\bar{Q}_{-1}\right)= -P+K_2\\
&&\left(Q_1-Q_{-1}\right) +  \left(\bar{Q}_1-\bar{Q}_{-1}\right)= 2iB\\
&&\left(Q_1-Q_{-1}\right) -  \left(\bar{Q}_1-\bar{Q}_{-1}\right)= -2iD 
\end{eqnarray}

The ground state $\ket{\Psi}$ of CFT Hamiltonian $H$ on the line is the only state in the Hilbert space invariant under the global conformal group and thus annihilated by all these generators. More generally, $\ket{\Psi}$ is annihilated by 
\begin{equation}
Q_n \ket{\Psi} = \bar{Q}_n \ket{\Psi} =0,~~~~\forall n\geq -1.
\end{equation}
Notice in particular that $\ket{\Psi}$ is also the ground state of 
\begin{equation}
Q_0 + \bar{Q}_0 = \int dx~ \frac{1+x^2}{2}  h(x) =  \frac 1 2 (H+K_1)
\end{equation}
While the Hamiltonian $H=\int dx~h(x)$ has a continuous spectrum, $Q_0 + \bar{Q}_0$ has the same discrete spectrum as $L_0+\bar{L}_0$ on the circle, which is given in terms of the scaling dimensions $\Delta_{\alpha} \equiv h_{\alpha}+\bar{h}_{\alpha}$ of the local scaling operators $\phi_{\alpha}(x)$ of the theory [similarly, while the spectrum of $P$ is continuous, the spectrum of $Q_0 - \bar{Q}_0 = \frac 1 2 (P+K_2)$ is discrete and given by  the conformal spins $s_{\alpha}\equiv h_{\alpha}-\bar{h}_{\alpha}$].

\section{1+1 free boson CFT on the real line}

In this appendix we specialize to the 1+1 dimensional free boson CFT on the real line discussed in this paper. We collect from \cite{CFTbook} a number of well-known fact that are used in the main text.

\subsection{Stress-energy tensor}

Consider the action
\begin{equation}
S \equiv \int dx~dt~ \frac{1}{2} \left((\partial_t\phi)^2 - (\partial_x \phi)^2\right)
\end{equation}
which leads to a stress-energy tensor
\begin{eqnarray}
T(x) &=& -2\pi ~:\partial \phi(x) \partial \phi(x):\\
\bar{T}(x) &=& -2\pi ~:\bar{\partial} \phi(x) \bar{\partial} \phi(x):
\end{eqnarray}
where $\partial \phi$ is a holomorphic (or right moving) field and $\bar{\partial}\phi$ is an antiholomorphic (or left moving) field,
\begin{eqnarray}
\partial \phi(x) &\equiv& \frac{1}{2}\left(\partial_x \phi(x) - i\partial_\tau \phi(x)\right)\\
 &=& \frac{1}{2}\left(\partial_x \phi(x) - \pi(x)\right),\\
\bar{\partial} \phi(x) &\equiv& \frac{1}{2}\left(\partial_x \phi(x) + i\partial_\tau \phi(x)\right)\\
&=&\frac{1}{2}\left(\partial_x \phi(x) + \pi(x)\right),
\end{eqnarray}
and where we have identified $\partial_t \phi(x)$ with $\pi(x)$. Therefore the normal-ordered energy and momentum densities are given by 
\begin{eqnarray}
h(x) &\equiv&   ~~: \frac{1}{2}\left[\pi(x)^2 + (\partial_x\phi(x))^2 \right]:~,\\
p(x) &\equiv&  ~ -:\pi(x)\partial_x\phi(x):~. 
\end{eqnarray}

\subsection{Generators of global conformal transformations}

Introducing now the annihilation operators $a(k)$ that diagonalize the Hamiltonian $H = \int dx~h(x) = \int dk~ |k|~ a(k)^{\dagger} a(k)$, namely 
\begin{eqnarray} \label{eqApp:aCFT}
a(k) &\equiv& \sqrt{\frac{|k|}{2}}\phi(k) + i  \sqrt{\frac{1}{2|k|}}\pi(k),
\end{eqnarray}
where $\phi(k) \equiv \frac{1}{\sqrt{2\pi}}\int dx~ e^{-ikx} \phi(x)$ and $\pi(k) \equiv \frac{1}{\sqrt{2\pi}}\int dx~ e^{-ikx} \pi(x)$ and $[a(k),a(q)^{\dagger}] = \delta(k-q)$, we can write the global conformal generators as
\begin{eqnarray}
H &=& \int dk~ |k|~ a(k)^{\dagger} a(k),\\
P &=& \int dk~ k~ a(k)^{\dagger} a(k),\\
B &=& i \int dk~ a(k)^{\dagger} \left(k\partial_k + \frac{1}{2}\right)a(k), \\
D &=& i\int dk~  a(k)^{\dagger} sgn(k)\left(k\partial_k + \frac{1}{2}\right)a(k),\\
K_1 &=& -\int dk~ a(k)^{\dagger}sgn(k) \left(k\partial^2_k +  \partial_k -\frac{1}{4k}\right)a(k),~~~~~~\\
K_2 &=& -\int dk~a(k)^{\dagger} \left(k\partial^2_k + \partial_k -\frac{1}{4k}\right)a(k).
\end{eqnarray}

\subsection{Primary operators and OPE's}

The primaries in this theory are $\mathbb{1}$, $\partial\phi$, $\bar\partial\phi$ and the so-called vertex operator $:e^{i\alpha \phi}:$, with conformal dimensions $(h, \bar h)$ being $(0,0)$, $(1,0)$, $(0,1)$ and $(\frac{\alpha^2}{8\pi},\frac{\alpha^2}{8\pi})$ respectively.

From the free boson action $S$ above, we can compute the correlator 
\begin{equation}
\langle \phi(x) \phi(y)\rangle=-\frac{1}{4\pi}\log((x-y)^2) + const.,
\end{equation}
by Gaussian integration. It follows that, 
\begin{equation}
\langle\partial\phi(x)\partial\phi(y)\rangle=-\frac{1}{4\pi}\frac{1}{(x-y)^2},
\end{equation}
from which we can read the operator product expansion (OPE) of $\partial\phi$ with itself,
\begin{equation}
\partial\phi(x)\partial\phi(y)\sim-\frac{\mathbb{1}}{4\pi}\frac{1}{(x-y)^2},
\end{equation}
where $\sim$ indicates that we neglect regular terms. The stress tensor is precisely the ignored constant regular term,
\begin{eqnarray}
T(x) &\equiv& -2\pi:\partial\phi(x)\partial\phi(x): \\
&\equiv& -2\pi \lim_{y\rightarrow x} \left(\partial\phi(x)\partial\phi(y) - \langle \partial \phi(x) \partial \phi(y)\rangle \right).
\end{eqnarray}

We can then apply Wick's theorem to obtain other OPEs. For example,
\begin{eqnarray}
T(x)\partial\phi(y)&=&-2\pi :\partial\phi(x)\partial(x):\partial\phi(y)\\
&\sim& \frac{\partial\phi(y)}{(x-y)^2} +\frac{\partial_y(\partial\phi(y))}{(x-y)}
\end{eqnarray}
and
\begin{eqnarray}
T(x)T(y)&=&4\pi^2 :\partial\phi(x)\partial\phi(x)::\partial\phi(y)\partial\phi(y):\\
&\sim& \frac{1/2}{(x-y)^4} + \frac{2T(y)}{(x-y)^2} +\frac{\partial_yT(y)}{(x-y)}.
\end{eqnarray}

All these expressions hold also for the cMERA smeared operators $\partial \phi^{\cM}(x) = V \partial\phi(x)V^{\dagger}$, $T^{\cM}(x) = V T(x)V^{\dagger}$, etc. They can be obtained either by applying the symplectic map $V$ on the CFT expressions or, equivalently, by defining normal ordering with respect to the cMERA annihilation operators $a^{\cM}(k)$ and applying Wick's theorem directly on the smeared operators.


\begin{thebibliography}{99}
\bibitem{Euclidean}
Right and left moving fields are the Lorentzian version of holomorphic and antiholomorphic fields in Euclidean CFT \cite{CFTbook}.
\bibitem{compareJutho} For general $x$ the conditions relating to dilations and boosts read $-i[D^{\cM},O^{\cM}_{\alpha}(x)] = \left(x\partial_x + \Delta_{\alpha}\right) O^{\cM}_{\alpha}(x)$  and $
-i[B^{\cM},O^{\cM}_{\alpha}(x)] = s_{\alpha} O^{\cM}_{\alpha}(x) -i~x[H^{\cM},O^{\cM}_{\alpha}(x)]$ for boosts. A version of the first condition was already mentioned in Ref. \cite{cMERA}. We note, however, that the discussion in Ref. \cite{cMERA} referred to the limit $\Lambda \rightarrow \infty$, whereas here we are concerned with the finite $\Lambda$ case.
\end{thebibliography}
\end{document}